\documentclass[transmag]{IEEEtran}
\usepackage{latexsym}
\usepackage{graphicx}
\usepackage{amsfonts,amssymb,amsmath}
\usepackage{hyperref}
\usepackage{xcolor}
\usepackage{cite}
\usepackage[utf8]{inputenc}
\usepackage[english]{babel}
\usepackage{algorithmic}
 \usepackage{algorithm}
 \usepackage{enumerate}

\usepackage{lipsum}
\def\BibTeX{{\rm B\kern-.05em{\sc i\kern-.025em b}\kern-.08em T\kern-.1667em\lower.7ex\hbox{E}\kern-.125emX}}
\begin{document}

\title{On Topology Optimization and Routing in Integrated Access and Backhaul Networks: A Genetic Algorithm-based Approach}

\author{Charitha Madapatha, Behrooz Makki,
\IEEEmembership{Senior Member, IEEE,}
\\ Ajmal Muhammad,
\IEEEmembership{Senior Member, IEEE,} Erik Dahlman, 
\\
Mohamed-Slim Alouini, 
\IEEEmembership{Fellow, IEEE}
and Tommy Svensson,
\IEEEmembership{Senior Member, IEEE.}
\thanks{Manuscript received xxxx; accepted November xxxx. Date of publication xxxx; date of current version xxxx. This work was supported in part by VINNOVA (Swedish Government Agency for Innovation Systems) within the VINN Excellence Center ChaseOn and in part by the European Commission through the H2020 project Hexa-X (Grant Agreement no. 101015956).
The work of C. Madapatha in this publication is part of his research work at Chalmers University of Technology.}
\thanks{C. Madapatha and T. Svensson are with the Department of Electrical Engineering, Chalmers University of Technology, 412 96 Gothenburg, Sweden (e-mail: charitha@chalmers.se;  tommy.svensson@chalmers.se).}
\thanks{B. Makki is with Ericsson research, Lindholmen, Sweden, 417 56 Göteborg, Sweden (e-mail: behrooz.makki@ericsson.com).}
\thanks{A. Muhammad is with Ericsson research, Kista, Sweden, P.O. Box 16440 (e-mail: {ajmal.muhammad}@ericsson.com).}
\thanks{E. Dahlman is with Ericsson research, Kista, Sweden, P.O. Box 16440 (e-mail: {erik.dahlman}@ericsson.com).}

\thanks{M.-S. Alouini is with the Computer, Electrical and Mathematical Science and Engineering, King Abdullah University of Science and Technology (KAUST), Thuwal 23955-6900, Saudi Arabia (e-mail: slim.alouini@kaust.edu.sa).}
}

\IEEEtitleabstractindextext{\begin{abstract}
In this paper, we study the problem of topology optimization and routing in integrated access and backhaul (IAB) networks, as one of the promising techniques for evolving 5G networks. We study the problem from different perspectives. We develop efficient genetic algorithm-based schemes for both IAB node placement and non-IAB backhaul link distribution, and evaluate the effect of routing on bypassing temporal blockages. Here, concentrating on millimeter wave-based communications, we study the service coverage probability, defined as the probability of the event that the user equipments’ (UEs) minimum rate requirements are satisfied. Moreover, we study the effect of different parameters such as the antenna gain, blockage, and tree foliage on the system performance. Finally, we summarize the recent Rel-16 as well as the upcoming Rel-17 3GPP discussions on routing in IAB networks, and discuss the main challenges for enabling mesh-based IAB networks. As we show, with a proper network topology, IAB is an attractive approach to enable the network densification required by 5G and beyond.

\end{abstract}

\begin{IEEEkeywords}
 Integrated access and backhaul, IAB, Genetic algorithm, Node selection, Topology optimization, \textcolor{black}{Densification},  Millimeter wave, (mmWave) communications, 3GPP, Stochastic geometry, Poisson point process, Coverage probability, Germ-grain model, Wireless backhaul, 5G NR,  Blockage, Relay, Routing, Tree foliage, Machine learning
\end{IEEEkeywords}

}

\maketitle
\section{INTRODUCTION}
Several reports have shown an exponential growth of demand on wireless communications, the trend which is expected to continue in the future \cite{eref1}. To cope with such demands, 5G and beyond networks propose various methods for capacity and spectral efficiency improvement. Here, one of the promising techniques is network densification, i.e., the deployment of many base stations (BSs) of different types such that there are more resource blocks per unit area \cite{eref2,eref3,eref4,eref5}.

The BSs need to be connected to the operators’ core network via a transport network, the problem which becomes challenging as the number of BSs increases. Such a transport network may be provided via wireless or wired connections. Wired (fiber) connections are typically used for transport closer to the core network and in the core network, where we need to handle aggregated traffic from multiple BSs.  Wireless connections, on the other hand, are used for backhaul transport in the radio access network (RAN) closer to the BSs. 

As reported in \cite{eref2}, the backhaul technology has large regional variations. However, on a global scale, wireless microwave technology has been a dominating media for the last few decades. Recently, there is an increase in fiber deployments attributed to geopolitical decisions and major governmental investments. Thus, going forward, it is expected that microwave and fiber will be two dominating backhaul technologies.

Fiber offers reliable connection with high peak data rates. However, 1) the deployment of fiber requires a noteworthy initial investment for trenching and installation, 2) may take a long installation time, 3) may be even not allowed in, e.g., metropolitan areas. Wireless backhaul using microwave is a well-established alternative to fiber, providing 10’s of Gbps in commercial deployments\footnote{Recent results demonstrate even more than 100 Gbps over MIMO backhaul links \cite{eref7}.}. Importantly, microwave is a scalable and economical backhaul technique that can meet the increasing requirements of 5G networks. Compared to fiber, wireless backhauling comes with significantly lower cost and time-to-market as well as higher flexibility, with no digging, no intrusion or disruption of infrastructure, and is possible to deploy in principle everywhere \cite{eref2}.

With the same reasoning and motivated by availability of massive bandwidth in millimeter wave (mmWave) spectrum/network densification, integrated access and backhaul (IAB) network has recently received considerable attention \cite{eref8,eref9,terabit_surv}.  With IAB, the goal is to provide flexible wireless backhauling using 3GPP new radio (NR) technology in international mobile telecommunications bands, and provide not only backhaul but also the existing cellular services in the same node and via the same hardware. This, in addition to creating more flexibility and reducing the time-to-market, is generally to reduce the cost for a wired backhaul, which in certain deployments could impose a large cost for the installation and operation of the BS. Importantly, 
\begin{itemize}
\item Internal evaluations at
Ericsson shows that,  even in the presence of dark fiber, the deployment of IAB network gives an opportunity to reduce the total cost of ownership in urban/suburban areas. This is partly because the same hardware can be used both for access and backhaul, i.e., less extra equipment is required \textcolor{black}{especially for in-band backhauling.} 
\item An integrated access/backhaul solution improves the possibilities for pooling of spectrum where it can be up to the operator to decide what spectrum resources to use for access and backhaul, rather than having this decided in an essentially static manner by spectrum regulators.
\end{itemize} 

\textcolor{black}{
In this way, IAB serves as a complement to microwave and fiber backhaul specially in dense urban and suburban deployments.}

\textcolor{black}{
Although IAB can in principle operate in every spectrum for which NR operation is specified, the focus of the 3GPP work on IAB has been on mmWave spectrum. This is intuitive because of the access to wide bandwidth in mmWave spectrum, while the existing LTE spectrum is very expensive to be used for backhauling. With a mmWave spectrum, however, blockage and tree foliage may be challenging, as they reduce the achievable rate significantly. Properly planned and optimized networks could reap higher performances, and save costs to network operators as they can avoid static blockages such as buildings/trees \cite{eref27}, \cite{eref28}. On the other hand, along with enabling traffic-based load balancing, routing can well compensate for temporal blockages, e.g., busses/trucks passing by. However, as the network size increases in dense areas, which is the main point of interest in IAB networks, deriving closed-form solutions for optimal network topology/routing becomes infeasible. }

\textcolor{black}{
It should be mentioned that the network deployment optimization can be done offline, and recalculated whenever there are substantial changes in the blocking situations, service rate requirements, and addition of new set of BSs.
Still, the optimization problem quickly becomes very large, thus motivating a potentially suboptimal machine learning approach, since an exhaustive search over all possible deployment options quickly becomes infeasible (see Section \ref{res_sec} for further details).
In such cases, machine learning techniques give effective (sub)optimal solutions with reasonable implementation complexity.} 
\subsection{Literature Review}
\textcolor{black}{
The performance of IAB networks have been studied from different perspectives. Particularly, \cite{eref10,eref11,eref12,eref13,eref14},  develop various resource allocation schemes, and  \cite{eref15}, \cite{eref16} study the effect of time/frequency division duplex based resource allocation on the throughput of IAB networks.  Moreover, \cite{eref17,eref18,eref19} utilize infinite Poisson point processes (PPPs) to evaluate the coverage probability of multi-hop IAB networks. Then, \cite{eref20,eref21} investigate the feasibility/challenges of mmWave-based IAB networks via end-to-end simulations. Also, \cite{eref22} and \cite{eref23} evaluate the potentials of using IAB in fixed wireless access and unmanned aerial vehicle-based communication setups, respectively. In \cite{eref24} and \cite{erefour}, we provide an overview of 3GPP Rel-16 discussions on IAB. Moreover, \cite{erefour} uses a FHPPP (FH: finite homogeneous), i.e., a PPP with a constant density and random distributions of the nodes in a finite region, to analyze the performance between the IAB and fiber-connected networks, and verify the robustness of IAB to various environmental effects. Finally, \cite{eref25} develops simulated annealing algorithms for joint scheduling and power allocation, and \cite{eref26} designs a joint precoder design and power allocation scheme maximizing the network sum rate.}

\textcolor{black}{
The problem of routing in IAB networks has been previously studied in the cases with different numbers of hops \cite{eref10,eref11,eref12,eref13,eref14,eref30,eref31,eref32}. Also, \cite{eref24} develops a cost-optimal node placement scheme, and \cite{eref33} proposes a joint node placement and resource allocation scheme maximizing the downlink sum rate of IAB networks. On the other hand, with different (non-IAB) network topologies/use-cases, various machine-learning based solutions have been previously proposed for topology optimization. For instance, \cite{gen4,gen5,gen6,gen7,gen8,gen9,gen10,gen11,gen12,gen13} develop deep reinforcement (DR) algorithm-based solutions for topology optimization of different network configurations. Deep Q-learning is used in \cite{gen1} to evaluate the cumulative transmission rate in vehicular networks. Spectrum allocation and access mode selection evaluations are considered in \cite{gen2}, while the potential of using $K$-means clustering algorithm to design ultra-reliable and low-latency wireless sensor networks is evaluated in \cite{gen3}. In addition, \cite{gen14} and \cite{gen15} use DR learning-based algorithms to solve the large-scale load balancing problem for ultra-dense networks. }

\textcolor{black}{Note that the mentioned works neither consider the non-IAB backhaul links distribution optimization nor the joint optimization of the non-IAB backhaul links distribution and the IAB nodes placement. This, although is of interest in practice, may be due to the fact that such optimization problems are NP-hard with a large search space. Therefore, one needs to design efficient algorithms which can find (semi)optimal solutions within a limited simulation period. Moreover, \cite{eref10}-\cite{gen15} concentrate on multi-hop communications, while the usefulness and challenges of meshed-based IAB have not \textcolor{black}{yet been} studied. Here, it is important to consider both the performance evaluations and the standardization issues, as meshed IAB has not \textcolor{black}{yet been} discussed in 3GPP 5G NR. These are the motivations for our work as presented in the following.}

\subsection{Contributions}

\textcolor{black}{
In this paper, we study the problem of topology optimization and routing in IAB networks. We study the problem from different points of views: }
\begin{itemize}

\item \textcolor{black}{We design effective genetic algorithm (GA)-based techniques not only for IAB node placement but also for dedicated non-IAB backhaul connection distribution. Here, concentrating on the characteristics of mmWave communications, we present the results for the cases with an FHPPP-based stochastic geometry model \cite{eref17}, \cite{eref29}. As the metric of interest, we consider the network service coverage probability which is defined as the probability of the event that the UEs’ minimum data rate requirements are satisfied.}  

\item \textcolor{black}{We study the effect of temporal blockages and routing on the coverage probability. In this way, one can avoid both the long-term and temporal blockages via topology optimization and routing, respectively. Also, the setup gives hints on the effectiveness of mesh-based communication in IAB networks, although it is not yet considered by 3GPP IAB standardization.} 

\item \textcolor{black}{We summarize the main 3GPP Rel-16 agreements as well as the upcoming Rel-17 discussions on routing, and highlight the main challenges which need to be solved before meshed IAB can be implemented.} 

\item \textcolor{black}{We study the effect of different parameters such as antenna gain, blockage and tree foliage on the system performance in both cases with well-planned and random network deployments.} 

\item \textcolor{black}{Finally, we compare the performance of the GA-based scheme with different state-of-the-art topology optimization methods. Also, we study the efficiency of the deployment optimization in the cases with constraint on the network topology, where the IAB nodes and the non-IAB backhaul links can not be freely deployed in every place. }
\end{itemize}

Compared to the related literature, e.g., \cite{eref27}-\cite{eref29}, we consider \textcolor{black}{more realistic} algorithms and network configurations. Moreover, our discussions on the effect of environmental parameters/deployment constraints on the system performance as well as the 3GPP agreements on IAB-based routing have not been presented before. Also, we optimize the IAB network for both node locations and non-IAB backhaul link placement independently, as well as jointly which further improves the coverage probability. We compare the performance of the proposed algorithms with different state-of-the-art schemes. These make our discussions and the conclusions completely different from those presented in the state-of-the-art works.

\textcolor{black}{As we show, machine learning techniques provide effective solutions for deployment optimization which can be easily adapted for different channel models, constraints and metrics of interest with no need for mathematical analysis. Moreover, compared to random deployment, deployment planning increases the coverage probability of the IAB networks significantly. On the other hand, with a well-planned network and for a broad range of blockage/tree foliage densities, the network can well handle these blockages with small routing updates. Finally, while the service coverage probability of the IAB network is slightly affected by stationary/temporal blockages in urban areas, for a broad range of parameter settings, the blockage is not problematic for well-planned routing-enabled IAB networks, in the sense that its impact on the coverage probability is negligible. On the other hand, high levels of tree foliage may reduce the coverage probability of the network in suburban areas; the problem which can be solved by proper deployment planning.} 
\section{IAB in 3GPP} \label{3gpp_sec}
IAB was introduced as part of Rel-16 of the 5G NR specification, with the specification finalized in fall 2020 \cite{3gpp_rp201756}. Currently, Rel-17 work item on IAB enhancements is going on, which is expected to finish in early 2022 \cite{3gpp_rp201293}.

The overall architecture for IAB is based on the \textcolor{black}{CU/DU split} of the gNB, introduced already in 3GPP Rel-15. With such architecture, a gNB consists of two functionally different parts with a standardized interface (referred to as the F1 interface) in between:

\begin{itemize}
    \item \textcolor{black}{A Centralized Unit (CU)} including the packet data convergence protocol (PDCP) and radio resource control (RRC) protocols,
    \item One or several \textcolor{black}{Distributed Units (DUs)} consisting of radio link control (RLC), medium access control (MAC), and physical layer protocols. 
\end{itemize}

\begin{figure}
\centerline{\includegraphics[width=3.5in]{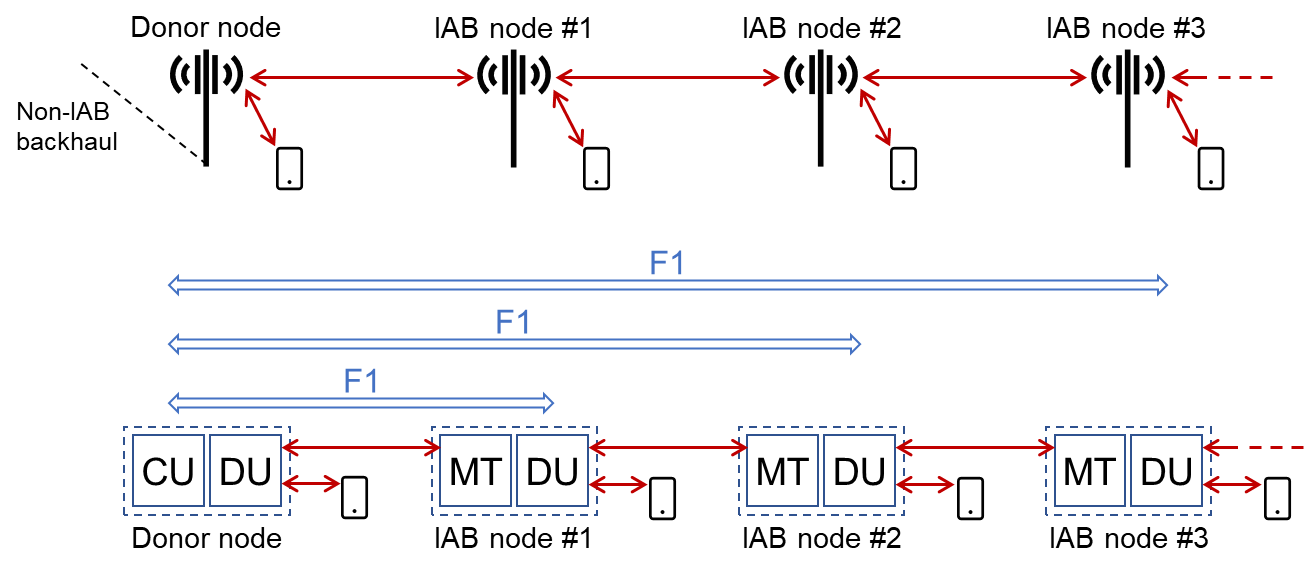}}
\caption{Types of network nodes.\label{types_nodes}}
\end{figure}

IAB specifies two types of network nodes (see also Fig. \ref{types_nodes}):

\begin{itemize}
    \item The IAB-Donor-node is the node consisting of CU and DU functionalities, and connects to the core network via non-IAB, for example fiber, backhaul.
    \item The IAB node \textcolor{black}{includes two modules}, namely, DU and mobile terminal (MT). IAB-DU serves UEs as well as, potentially, \textcolor{black}{downstream} IAB nodes in case of multi-hop wireless backhauling. At its other side, an IAB-MT is the unit \textcolor{black}{that connects an IAB node with} the DU of the parent\textcolor{black}{/upstream node.}
\end{itemize}

The IAB architecture is thus based on a hierarchical or, at least, a-cyclical structure where it is well-defined if a certain node is “above” or “below” a certain other node and where information flows in well-defined down-stream and up-stream directions. The possibility for a more mesh-like structure with no well-defined hierarchy was briefly discussed during the initial phase of the 3GPP work on IAB. However, \textcolor{black}{majority of the companies discard the idea owing to its complexity and no clear benefits. }

\begin{figure}
\centerline{\includegraphics[width=3.5in]{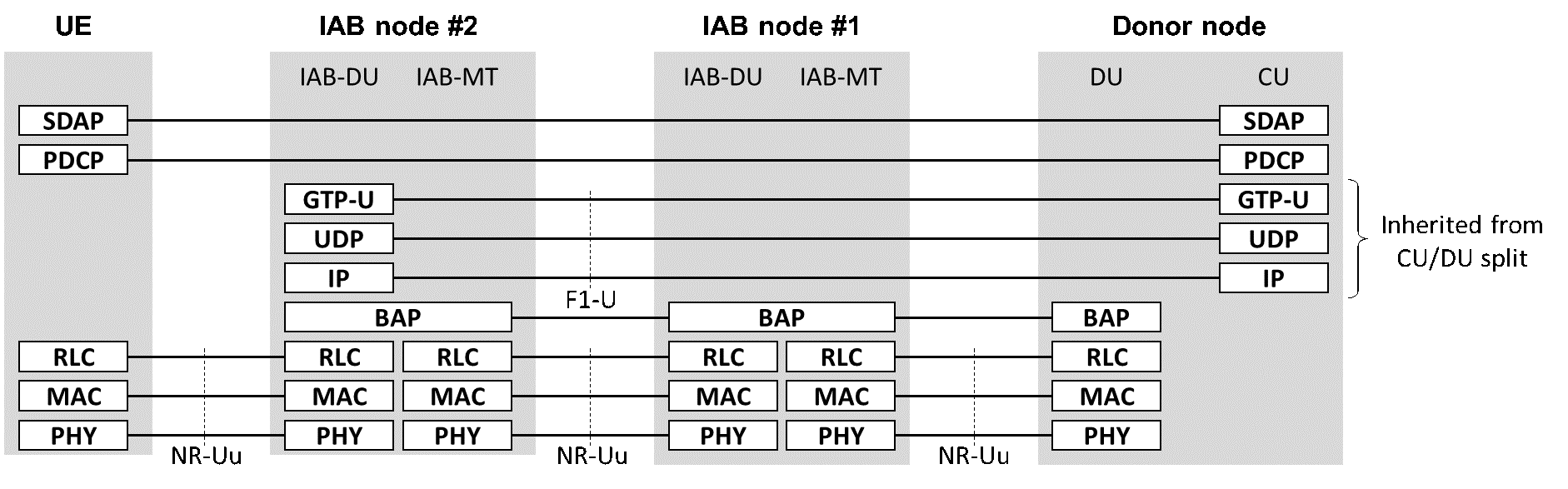}}
\caption{IAB protocol stack of the user-plane.\label{ustack}}
\end{figure}

\textcolor{black}{When it comes to the connectivity with the parent node, the IAB-MT connects to the DU of its parent node essentially as a normal UE.} The \textcolor{black}{Uu interface, i.e., the link} between the parent node DU and the MT of the IAB node then provides the lower-layer functionality \textcolor{black}{and relays the F1 messages between} the donor-node CU and the IAB-node DU.  
The specification of the F1 interface only defines the higher-layer protocols, for example, the signaling messages between the CU and DU, but is agnostic to the lower-layer \textcolor{black}{(i.e., transport network layer)} protocols. With IAB, the NR radio-access technology (the RLC, MAC, and physical layer protocols) together with some IAB-specific protocols, provides the lower-layer functionality on top of which the F1 interface is implemented. See Fig. \ref{ustack} showing the user-plane protocols of a multi-hop IAB network (the control plane has a similar structure).

\subsection{Backhaul Adaptation Protocol}
\begin{figure}
\centerline{\includegraphics[width=3.5in]{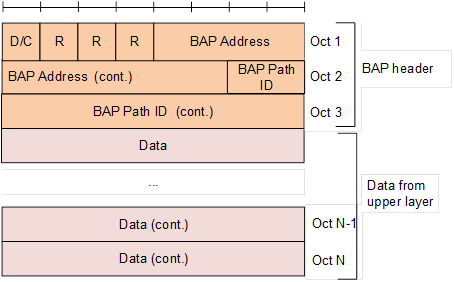}}
\caption{
Structure for the BAP header.\label{bapfig}}
\end{figure}

Backhaul adaptation protocol (BAP) is a new IAB-specific protocol responsible for routing and bearer mapping of packets in the IAB network. More specifically, the BAP layer is responsible for forwarding of the packets in the intermediate nodes/hops between the IAB-donor-DU and the access IAB-node. For the downstream traffic, the BAP layer of the IAB-Donor-DU will add a BAP header to packets received from the upper layer. Similarly, for the upstream traffic, the BAP layer of the access IAB-node will add a BAP header to the upper layer packets. Figure \ref{bapfig} shows the structure for the BAP header, which contains a 10-bit BAP address field and a 10-bit BAP path ID field apart from 1-bit flag and 3 reserve bits for future use. Note that 3GPP specifications use the BAP Routing ID as a cover term for BAP address and BAP path ID fields. The purpose of the BAP address field is to carry the address of the destination IAB-node, while the Path ID field contains the path identity to be used for traversing the packets towards the destination IAB-node. This latter field is important for situations where multiple paths are configured for an IAB-node to improve network robustness/resilience and achieve load balancing by transporting a part of the traffic via each path towards the IAB-node.

\begin{figure}
\centerline{\includegraphics[width=3.5in]{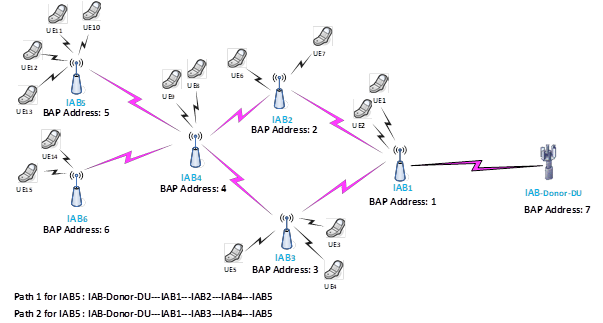}}
\caption{
Example of routing in IAB network.\label{iabnetwork}}
\end{figure}

To illustrate the above concept, figure \ref{iabnetwork} shows an example topology for IAB network where two paths (i.e., Path 1 and Path 2) have been configured for IAB-node 5 by the IAB-donor-CU. This means that the routing tables in the BAP layer of all the intermediate IAB-nodes (i.e., IAB1, IAB2, IAB3, etc.) are properly configured with next-hop link information for all the BAP addresses and BAP path IDs carried in the packets BAP header that these nodes will route in the network. Furthermore, the IAB-donor-DU will have mapping rules (configured by the donor-CU) how to select the BAP address and BAP path ID fields for packets from the upper layer based on the information in the IP address fields (i.e., DS/DSCP) of the F1-AP signaling.

Suppose the IAB-donor-DU receives a packet with IP address fields marked with information that is mapped to BAP address 5 and BAP path ID 1, the donor-DU will add a BAP header with proper field values (i.e., address 5 and path ID 1) and will forward the packet to IAB2. Once IAB2 receives the packet, the node will examine the BAP header of the packet and based on the BAP address (carried in the packet) and its routing table information will transmit the packet towards IAB4. Similarly, IAB4 will route the packet to IAB5, where the IAB5 upon examining the BAP header field of the packet will notice that the packet is destined for it. Hence, IAB4 will remove the BAP header before delivering the packet to its upper layer for further processing.

In another scenario, if IAB1 receives a packet (from IAB-donor-DU) with BAP header containing BAP address 5 and path ID 2, IAB1 will forward the packet towards IAB3 instead of IAB2, and so on IAB3 will forward the packet to IAB4. When it comes to the upstream traffic, the BAP layer of IAB5 will add a BAP header containing IAB-donor-DU BAP address and appropriate path ID (either path ID 1 or path ID 2 based on the configuration information) to packets received from the upper layer. Next, IAB5 will forward the packets to IAB4, which will be further forwarded by IAB4 either to IAB1 or IAB2 depending on the path ID field value carried in the packets BAP headers. Once the packets reach IAB-donor-DU, the DU will remove the BAP header before delivering the packets to the upper layer for subsequent processing.   

\subsection{IAB extensions in 3GPP release 17}

\begin{figure}
\centerline{\includegraphics[width=3.5in]{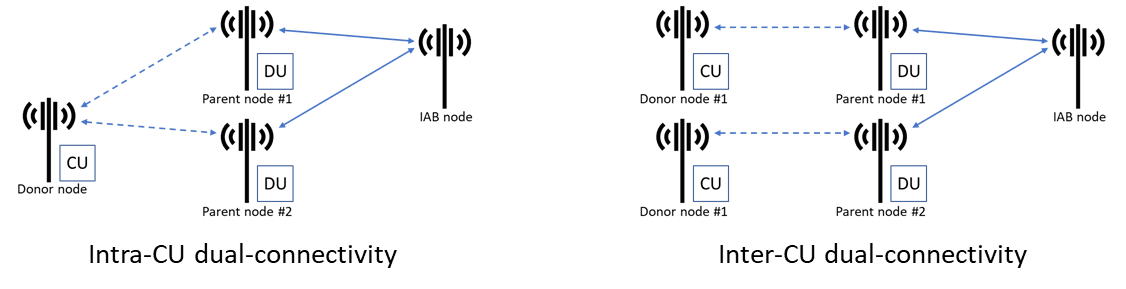}}
\caption{
Intra-CU dual connectivity vs inter-CU dual-connectivity \label{intracu}}
\end{figure}

3GPP is considering further extensions and enhancements to IAB as part of NR Rel-17. One topic for Rel-17 is to look further into the support for dual-connectivity scenarios for IAB. For the regular network-to-device link, dual-connectivity, supported for both NR and 4G/LTE, implies that a device has established a link to multiple cells operating on different carrier frequencies. In the context of IAB, dual-connectivity like-wise implies that an \textcolor{black}{IAB-DU is connected to multiple parent nodes via its collocated IAB-MT.} Such IAB dual connectivity can be either intra-CU, that is, \textcolor{black}{the same donor node is serving both parent nodes. Alternatively, dual-connectivity can be inter-CU, that is, there are multiple IAB-Donor-nodes (see Fig. \ref{intracu}). Clearly, the inter-CU dual-connectivity has a larger impact on the IAB network in terms of specification work and complexity.}

IAB dual-connectivity is envisioned to provide higher reliability due to an additional redundancy in the wireless backhaul. It may also enable additional possibilities for load-balancing within the wireless backhaul, i.e., the possibility to more dynamically route data via different paths depending on the instantaneous load \textcolor{black}{conditions} on different links.

\section{System Model}\label{system_model}
This section presents the system model, including the channel
model, the considered UE association rule as well as
the achievable data rates in the backhaul and access links. Table \ref{tab1x} summarizes the parameters used in
the analysis.
\begin{figure*}[t]
\centerline{\includegraphics[width=7.5in]{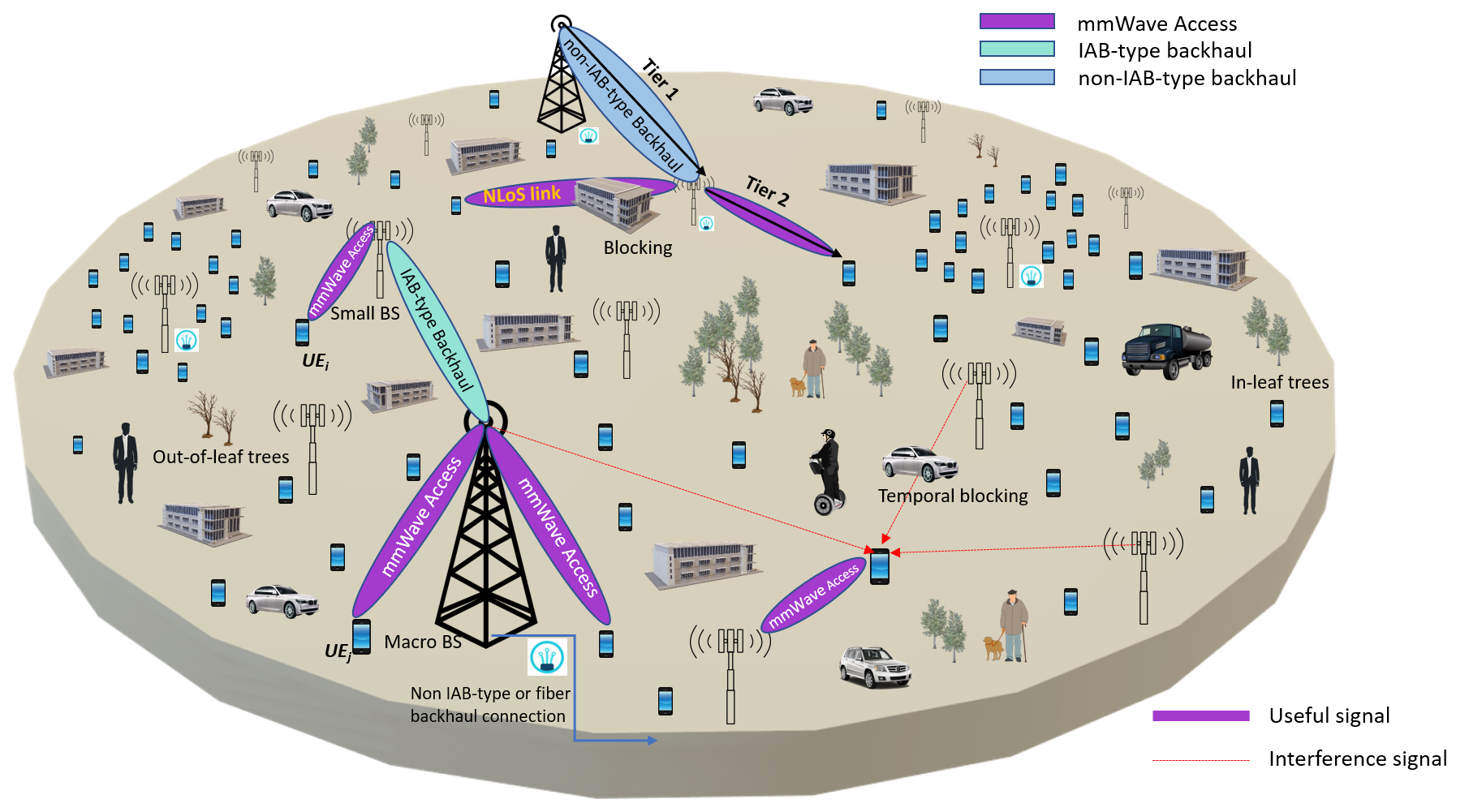}}
\caption{Schematic of the considered \textcolor{black}{ IAB network}. \textcolor{black}{A majority of the SBSs rely on IAB for backhauling. A small fraction of the SBSs, however, may have non-IAB type backhaul where such a backhauling can be provided by fiber or wirelessly.}\label{systemmodelfig}}
\end{figure*}
Consider a dense urban area with a two-tier heterogeneous
network (HetNet), i.e., a two-hop IAB network, where multiple MBSs (M: macro) and  SBSs (S: small) serve the UEs (see Fig. \ref{systemmodelfig}). In this way, following the 3GPP definitions (see Section \ref{3gpp_sec}), the MBSs and the SBSs represent the donor and the child IABs, respectively, and throughout the paper we may use the terminologies MBS/SBS and donor IAB/IAB interchangeably. With the IAB setup, both the MBSs and the SBSs are used for both access and backhaul. However, the donor IABs, i.e., the MBSs, are \textcolor{black}{non-IAB backhauled to the core network.}

\textcolor{black}{\textbf{Note.} In practice, a majority of the SBSs receive IAB-type backhaul from the MBSs wirelessly. However, a fraction of the SBSs may have access to non-IAB type dedicated backhaul connections, where such a backhaul can be provided either by fiber or a wireless radio link operating on a different frequency than the IAB network (see Fig. \ref{systemmodelfig}). In this paper, along with optimizing the SBSs’ locations, one of our goals is to determine the proper nodes with non-IAB type backhauling such that the coverage probability is maximized. In terms of performance evaluation, our analysis does not depend on if these non-IAB backhaul links are provided by fiber or wirelessly. In practice, however, the network performance may depend on the type of such non-IAB links; Wireless radio backhaul link is quite flexible, and one can provide an SBS with wireless non-IAB backhaul as long as there is a strong LoS connection between the SBS and an MBS. Fiber connections, on the other hand, may be available in specific areas, and there may be low flexibility in fiber distribution between the nodes. In summary, depending on the type of the non-IAB backhaul links, to the SBSs, our non-IAB backhaul link distribution method, presented in Algorithm 1, may give an ultimate opportunistic potential of the network performance.
}

\textcolor{black}{Note that the consideration of two-hop network, e.g., \cite{eref15, eref16, eref17}, \cite{erefour}, \cite{g1,g2}, in our work is motivated by the fact that, although the 3GPP standardization does
not limit the possible number of hops, as also reported in, e.g., \textcolor{black}{\cite{eref24}, \cite{gensys2,gensys3,gensys4}}, traffic aggregation in the backhaul links and end-to-end 
latency become challenging as the number of hops increases. Moreover, because IAB is of most interest in dense metropolitan areas with already existing limited number of fiber links, in most cases appropriate coverage can be provided with a maximum of two hops (For instance, see the Ericsson simulation results in the London area [\cite{g3}, Fig. 3] where the required QoS is satisfied mostly by only a single backhaul hop).}

\textcolor{black}{We} model the IAB network using \textcolor{black}{an FHPPP based random distribution of the nodes in a finite region \cite{eref17,eref18,eref19,eref26}.} Particularly, without topology optimization, which we use as the benchmark to evaluate the performance of planned networks, the FHPPPs $\chi_{\text{M}}, \chi_{\text{S}}, \chi_{\text{U}}$ with densities, $\lambda_{\text{M}}, \lambda_{\text{S}}$ and $\lambda_{\text{U}}$, respectively, are
used to model the spatial distributions of the MBSs, the SBSs and the UEs, respectively. With our topology optimization, however, while the number of nodes are still determined based on some random process, the locations of the SBSs as well as the location of the \textcolor{black}{non-IAB backhaul} connections to a fraction of the SBSs are optimized, in terms of network service coverage probability \textcolor{black}{(see Section \ref{res_sec} for the details). }

\begin{table*}[t]
\centering
\caption{The Definition of the Parameters.}
\label{table}
\setlength{\tabcolsep}{3pt}
\begin{tabular}{|p{40pt}|p{160pt}|p{40pt}|p{160pt}|}
\hline
Parameter& 
Definition&
Parameter&
Definition\\
\hline
$\chi_{\text{M}}$ & FHPPP of the MBSs & $\chi_{\text{U}}$ & FHPPP of the UEs \\
$\chi_{\text{S}}$& FHPPP of the SBSs&$\chi_{\text{bl}}$& FHPPP of the blockings\\
 $\chi_{\text{T}}$& FHPPP of the tree lines&$\lambda_{\text{U}}$& UEs density\\
 $\lambda_{\text{M}}$& MBSs density& $\lambda_{\text{S}}$& SBSs density\\
 $\lambda_{\text{bl}}$& Blocking density&$\lambda_{\text{T}}$& Tree density \\
 $\rho$&Service coverage probability&$d$&Vegetation depth\\
 $\theta$& Orientation of the blocking wall&$\varphi$&Angle between transmitter and receiver\\
 $D$&Circular disk&$R$&Radius of the disk\\
 $P_{\text{t}}$&Transmission power&$P_{\text{r}}$&Received power\\
 $h$&Fading coeficient&$G$&Antenna gain\\
 ${\gamma}_{(1\text{m})}$&Reference path loss at 1 meter distance&$\gamma$&Propagation path loss\\
 $x$&Location of the node&$r$&Propagation distance  between  the  nodes\\
 $\alpha$&Path  loss  exponent&$N$&Number of connected UEs\\
 $f_{\text{c}}$&Carrier frequency&$R_{\text{th}}$& Minimum data rate threshold\\
 $w_{\text{u}}$&Associated cell&$w_{\text{s}}$&Associated BS in backhaul link\\
 $N_\text{f}$&Number of \textcolor{black}{non-IAB backhauled} SBSs&$\kappa$&Tree foliage loss\\
 $N_\text{s}$&Number of SBSs&\textcolor{black}{$S_c$}&\textcolor{black}{Number of possible solution checkings}\\
 $l_\text{T}$&Tree line length&$l_\text{h}$&Hop length\\
 $\lambda_{\text{temp}}$&Temporal blocking density&$K$&Number of random sets in the GA\\
 $J$&Number of sets around the Queen in GA&$N_{\text{it}}$&Number of iterations in the GA\\
 $B$& Bandwidth & $\psi$ & Percentage of bandwidth resources on backhaul\\
\hline
\end{tabular}
\label{tab1x}
\end{table*}

In our setup, in-band communication is considered where
both the access and backhaul links share and operate in the same mmWave spectrum band. This is motivated by the fact that in-band communication gives better flexibility for resource allocation, at the cost of coordination complexity. For simplicity, assume the network to be distributed over a circular disk $D$. However, the model can be well applied on every arbitrary region $D$.

For the blockage, we use the well-known germ grain model described in  \cite[Chapter 14]{ref5}, which provides accurate blind spot prediction, compared to stochastic models that assume independant blocking. Particularly, the model takes induced blocking correlation into account and, thus, suits well for environments with large obstacles. Here, an FHPPP $\chi_{\text{bl}}$ models the blockage distribution in the area $D$ with $\lambda_{\text{bl}}$ denoting the density. The blockings are considered to be walls of length $l_{\text{bl}}$ and independantly and identically distributed orientation $\theta_{\text{bl}}$. Later, we  use a GA-based  approach  to  optimize  the SBSs locations inside the region $D$ to preferably avoid blockage in the backhaul links.

Following the state-of-the-art mmWave channel model, e.g., \cite{ref1}, the received power at each node can be expressed as
\begin{equation}
    P_{\text{r}}=P_{\text{t}}h_{\text{t,r}}G_{\text{t,r}}{\gamma}_{(1\text{m})}\gamma_{\text{t,r}}\left|\left|x_{\text{t}}-x_{\text{r}}\right|\right|^{-1}\kappa_{\text{t,r}}.
    \label{eqa1}
\end{equation}

Here, $P_{\text{t}}$ represents the transmit power in each link, and $h_{\text{t,r}}$ denotes the independant small-scale fading in individual links. Particularly, in our study Rayleigh fading is considered for small-scale fading. Thereby, $G_{\text{t,r}}$ denotes the combined antenna gain of the transmitter and receiver in the link, $\gamma_{\text{t,r}}$ is the propagation path loss, and ${\gamma}_{(1m)}$ is the reference path loss at one meter distance while $\kappa_{\text{t,r}}$ is the tree foliage loss. 

The total path loss, in dB, is characterized according to the 5GCM UMa close-in model described in \cite{ref3}. Here, the path loss is characterized by
\begin{equation}
\text{PL}=32.4+10\log_{10}(r)^\alpha+20\log_{10}(f_\text{c}),
\end{equation}
where $f_c$ is the carrier frequency, $r$ is the propagation distance between the nodes, and $\alpha$ is the path loss exponent. 
Depending on the blockage, line-of-sight (LoS) and NLoS (N: Non) links are affected by different path loss exponents. The propagation loss of the path loss model is given by 
\begin{equation}
\gamma_{\text{t,r}} = \begin{cases} r^{\alpha_{\text{L}}},&\text{if LoS,} \\ r^{\alpha _{{{\text{N}}}}},&\text{if NLoS,} \end{cases}
\end{equation}
where  $\alpha_\text{L}$ and $\alpha_\text{N}$ denote path loss exponents for \textcolor{black}{the} LoS and NLoS \textcolor{black}{scenarios}, respectively.

5G and beyond systems are equipped with large antenna arrays which are used to minimize the propagation loss. We use the sectored-pattern antenna array model to characterize the beam pattern and antenna gain, which is given by

\textcolor{black}{
\begin{equation}G_{\text{t,r} }(\varphi) = \begin{cases} G_0&\frac{-\theta_{\text{HPBW}}}2\leq\varphi\leq\frac{\theta_{\text{HPBW}}}2 \\ g(\varphi)&\text{otherwise.} \end{cases}
\end{equation}}
Here, $\varphi$ represents the angle between the transmit and receive antennas. Furthermore, $\theta_\text{{HPBW}}$ is the half power beamwidth, and $G_0$ denotes the main lobe gain of the antenna while $g(\varphi)$ is the side lobe gain \cite{ref1}. Finally, as in, e.g., \cite{eref17,eref19,erefour}, \textcolor{black}{for tractability} we assume that the UE antenna gain to be 0 dB due to its omni-directional beam pattern, \textcolor{black}{although UE beamforming in mmWave is an interesting future work to incorporate}.

\textcolor{black}{Unless otherwise stated and in harmony} with, e.g.,\cite{eref17,eref18,eref19},\cite{erefour}, we assume that the backhaul links are noise-limited. This assumption, which has been verified in \cite{erefour}, is motivated by the high beamforming capacity in the inter IAB backhaul links and the fact that simultaneous transmission/reception is not considered in our setup. \textcolor{black}{Then, Section \ref{res_sec} validates this assumption, and we verify} the effect of the backhaul interference on the coverage probability. Also, the inter-UE interference is neglected with the assumption of sufficient isolation and low power of the UEs \cite{eref22}. Particularly, the interference model focuses on the aggregated interference on the access links, caused by the neighbouring interferers, which for UE ${u}$ is expressed as    
\begin{equation} I_{{\rm {\textit{u}}}}= \sum \limits _{ {\textit{$\mathbf{i,u}$}\in \chi _{i,u}\setminus \{{\mathbf{w}}_{u}\}}}{P_{i}}  h_{i,u} G_{i,u}{\gamma}_{(1\text{m})} \gamma_{{x_i,x_u}}\|{\mathbf{x_\textit{i}-x_\textit{u}}}\|^{-1},
\label{inter}
\end{equation}
where $i$ denotes the set of BSs excluding the associated BS $w_\text{u}$ of user $u$. Also, for SBS $s$, the aggregated interference on the backhaul links is given by
\begin{align} I_{{\rm {\textit{s}}}}= \sum \limits _{ {\textit{$\mathbf{j,s}$}\in \chi_{j,s}\setminus \{{\mathbf{w}}_{s}\}}}{P_{j}}  h_{j,s} G_{j,s}{\gamma}_{(1\text{m})} \gamma_{{x_j,x_s}}\|{\mathbf{x_\textit{j}-x_\textit{s}}}\|^{-1},
\label{intersbs}
\end{align}

where $j$ denotes the set of transmitting BSs excluding the associated BS $w_\text{s}$ of SBS $s$.

We use an FHPPP denoted by $\chi_{\text{T}}$ with density $\lambda_{\text{T}}$ to model the spatial distribution of the tree lines of length $l_\text{T}$ \cite{gensys6}. The tree foliage loss is estimated using the Fitted International Telecommunication Union-Radio (FITU-R) tree foliage model \cite[Chapter 7]{gensys7}. The model is well known for its applicability in cases with non-uniform vegetation and frequency dependancy within 10-40 GHz range. Particularly, considering both in-leaf and out-of-leaf, vegetation states, the tree foliage loss in \eqref{eqa1} is expressed as  
\begin{equation}
   \kappa=\left\{\begin{array}{l}0.39f_{\text{c}}^{0.39}d^{0.25},\;\text{in-leaf}\\0.37f_{\text{c}}^{0.18}d^{0.59},\;\text{out-of-leaf,}\;\end{array}\right.
   \label{eqvege}
\end{equation}
where $d$ is the vegetation depth measured in \textcolor{black}{meter}.

In our setup, each UE has the ability to be connected to either an MBS or an SBS depending on the maximum average received power. Let $a_{u}\in \{0,1\}$ be a binary variable indicating the association with 1, while 0 representing the opposite. Thus, for the access links 
\begin{align} {a_u} = \begin{cases} 1 & \text {if}~P_{i} G_{z,x}h_{z,x}{\gamma}_{(1m)} \gamma_{\text{z,x}}(\|{\mathbf{z}}-{\mathbf{x}}\|)^{-1} \\&\qquad \quad ~~\geq P_{j}G_{j}h_{z,y}{\gamma}_{(1m)} \gamma_{\text{z,y}}(\|{\mathbf{z}}-{\mathbf{y}}\|)^{-1}, \\&\qquad \qquad \quad \forall ~{\mathbf{y}}\in \chi _{j}, j\in \{{\mathrm{ m}},{\mathrm{ s}}\}|{\mathbf{x}}\in \chi _{i}, \\ 0,& \text {otherwise,} \end{cases}\end{align} \label{eq:9}
where \textcolor{black}{$i$, $j$ denote the BS indices, i.e., MBS or SBS.} As in \eqref{eq:9} for each UE ${u}$, the association binary variable $a_u$ becomes 1 for the cell giving the maximum received power at the UE, while for all other cells it is 0, as the UE can only be connected to one IAB node. 

Since the MBSs and the SBSs have large antenna arrays and can beamform towards the desired direction, the antenna gain over the backhaul links can be assumed to be the same, and backhaul link association can be well determined  based on the minimum path loss rule, i.e., by
\begin{align}
{a_{b,m}} = \begin{cases} 1 & \text {if}~\gamma_{{\rm {b}}_{\mathrm{ m}}}(\|{\mathbf{z}}-{\mathbf{x}}\|)^{-1}\!\geq \! \gamma_{{\rm {b}}_{\mathrm{ m}}}(\|{\mathbf{z}}\!-\!{\mathbf{y}}\|)^{-1}, \\ &\qquad \qquad \qquad \qquad \qquad \forall ~{\mathbf{y}}\in \chi_{\mathrm{ m}}|{\mathbf{x}}\in \chi _{\mathrm{ m}}, \\ 0,& \text {otherwise.} \end{cases} \end{align}

\textcolor{black}{For resource allocation, on the other hand, the mmWave spectrum available is partitioned \textcolor{black}{into} the access and backhaul links such that} 

\begin{equation}\begin{aligned}
    \begin{cases}
    B_\text{Backhaul}=\psi B,\\
    B_\text{Access}=(1-\psi)B,
\end{cases}
\end{aligned}
  \label{eq:11}  
\end{equation}

\textcolor{black}{In practice, along with the MBSs which are} \textcolor{black}{non-IAB backhaul}-connected, a portion of the SBSs may have \textcolor{black}{dedicated non-IAB backhaul connections}, resulting in a hybrid IAB network. Therefore, in our deployment, some of the SBSs are \textcolor{black}{IAB backhauled wirelessly} and the others are \textcolor{black}{connected to dedicated non-IAB backhaul links.}

Let us initially concentrate on the IAB-type backhauled SBSs. Also, let, $B_\text{backhaul}$ and $B_\text{access}$ denote the backhaul and the access bandwidths, respectively, while total bandwidth is $B=B_\text{backhaul}+B_\text{access}$. The bandwidth allocated for each \textcolor{black}{IAB-type wirelessly backhauled} SBS, namely, child IAB, by the MBS, i.e., IAB donor, is proportional to its load and the number of UEs in the access link. The resource allocation is determined based on the instantaneous load where each IAB-type backhauled SBS informs its current load to the associated MBS each time.  Thus, the backhaul-related bandwidth for the $j$-th IAB node, if it does not have \textcolor{black}{dedicated non-IAB backhaul} connection, is given by  
\textcolor{black}{ 
\begin{equation}
{B_{\text{{backhaul}}{,j}}}=\frac{\psi BN_j}{{\displaystyle\sum_{\forall\;j}}N_j},{     \forall j},
\end{equation}}
\textcolor{black}{where $N_j$ denotes the number of UEs connected to the $j$-th IAB-type backhauled node and $\psi\in [0,1]$ is the fraction of the bandwidth resources on backhauling. \textcolor{black}{Therefore, the bandwidth allocated to the $j$-th IAB-type backhauled node is proportional to the ratio between its load, and the total load of its connected IAB donor.} Meanwhile, the access spectrum is equally shared among the connected UEs at the IAB node according to
\begin{equation}
{B_{\text{{access}},u}}=\frac{(1-\psi)B}{{\displaystyle\sum_{\forall\;u}}N_{j,u}}, \forall u,
\end{equation}
where  $u$ denotes the UEs indices, and $j$ represents each IAB-type backhauled node. Moreover, $N_{j,u}$ is the number of UEs connected to the $j$-th IAB-type backhauled node to which UE $u$ is connected.} Finally, the signal-to-interference-plus-noise ratio (SINR) is obtained in accordance with \eqref{inter} by \begin{align}
\text{SINR}=P_{\text{r}}/(I_{{u}}+\sigma^2),
\end{align}
where $\sigma^2$ is the noise power.

\textcolor{black}{With our setup, the network may have three forms of access connections, i.e., MBS-UE, \textcolor{black}{IAB-type backhauled} SBS-UE, \textcolor{black}{non-IAB backhauled} SBS-UE, and the individual data rates will
behave according to the form in which the UE’s connection has
been established. Particularly, the rates experienced by the UEs in access links that are connected to MBSs or to the IAB type-backhauled SBSs are given by }
\begin{equation}
 {R_{{u}}} =   \begin{cases}\frac{(1-\psi)B}{{}N_m}\log(1+\text{SINR}(x_{u})), ~\text { if }{\mathbf{w}}_\text{u}\in \chi _{\mathrm{ m}},\\ \min \bigg (\frac{(1-\psi)BN}{{\displaystyle\sum_{\forall\;u}}N_{j,u}}\log(1+\text{SINR}(x_{u})), \\ \qquad  \frac{\psi BN}{{\displaystyle\sum_{\forall\;j}}N_j}\log(1+\text{SINR}(x_\text{b}))\bigg ), \text {if }{\mathbf{w}}_\text{u}\in \chi _{\mathrm{ s}}, \end{cases}
 \label{eq:14}
\end{equation}
where $j$ represents each IAB-type backhauled SBS connected to the MBS.  Then, $m$ gives the associated MBS, $s$ denotes the SBS, and  $u$ represents the UEs' indices. \textcolor{black}{Unlike an MBS} which shares some of its bandwidth with \textcolor{black}{IAB-type backhauled SBSs, a non-IAB backhauled SBS} has a bandwidth of $B$ for access, and does not need \textcolor{black}{to share its bandwidth for backhauling}. Thus, the UEs connected to a \textcolor{black}{non-IAB backhauled SBS} experience the rate given by
\begin{equation}
{R_{u}} =  \frac{B}{{\displaystyle}N_\text{u}}\log(1+\text{SINR}(x_\text{u})) , \text {if }{\mathbf{w}}_\text{u}\in \chi _{\mathrm{ s},} 
\label{eq:fibsbs}
\end{equation}
where $N_u$ denotes the total number of UEs connected to the \textcolor{black}{non-IAB backhauled} SBS of which the considered UE is associated. Depending on the associated cell, there are three possible cases for the data rate of the UEs. First is the case when the UEs are connected to the MBSs, i.e., IAB donor, as denoted by $w_{\text{u}} \in \chi _{\mathrm{ m}}$ in \eqref{eq:14}. Since the MBSs have \textcolor{black}{non-IAB backhaul} connection, the rate will only depend on the access bandwidth available at the UE. In the second case, the UEs are connected to the IAB-type backhauled SBSs, as denoted by $w_{\text{u}} \in \chi _{\mathrm{s}}$ in \eqref{eq:14}. Here, the SBSs have shared backhaul bandwidth from the IAB-Donor-nodes  i.e., MBSs, and thus the UEs data rates depend on the backhaul rate of the connected IAB-type backhauled SBS as well. Thus, in this case the UE is bound\textcolor{black}{ed} to get the minimum between backhaul and access rate. Then, the third case is when the UEs are associates with the \textcolor{black}{non-IAB backhauled SBSs} as denoted in \eqref{eq:fibsbs}. Unlike in the previous case, here the SBSs have full bandwidth $B$ which is not shared with backhauling.

In the following, we present the GA-based schemes to optimize the locations of the SBSs as well as the \textcolor{black}{non-IAB backhaul link} distribution to a fraction of the SBSs such that the network service coverage probability is maximized.

\section{Proposed Algorithm}

In general, Rel-16 IAB network supports NLoS backhauling. However, the performance of the IAB networks is considerably affected by the quality of the backhaul links, where, if possible, it is preferred to have IAB-IAB channels with strong LoS signal strength. Also, in hybrid networks where a fraction of the SBS nodes may be \textcolor{black}{backhauled via dedicated non-IAB backhaul links}, it is important to obtain the set of SBSs that are critical to be \textcolor{black}{non-IAB backhaul}-connected for optimal performance. However, depending on the network size, it may be difficult to obtain the appropriate location of the SBSs and/or the \textcolor{black}{non-IAB backhaul link} placement scheme for SBSs analytically.

For instance, with $N_{\text{s}}$ SBSs and a budget of having $N_{\text{f}}$ \textcolor{black}{non-IAB backhauled} SBSs, there are $\binom{N_{\text{s}}}{N_{\text{f}}}$ possible combinations of \textcolor{black}{non-IAB backhauled} SBS selections. Therefore, the optimal set of SBSs suitable for \textcolor{black}{non-IAB backhaul link} placement can indeed be obtained via exhaustive search for the cases with few SBSs. However, as the network size increases, it is not feasible to search over all possible solutions. The problem becomes even
more challenging with determining the optimal locations of the
SBSs as they can be distributed in the whole network area. Thus, it is important to design efficient algorithms to obtain the (sub)optimal SBS locations as well as \textcolor{black}{dedicated non-IAB backhaul link} placement with low complexity. 

With this background, the state-of-the-art works mainly concentrate on either modeling the network by placing the BSs on a grid or distribute them randomly based on stochastic geometry models. However, none of these models are accurate, as they give an optimistic or a pessimistic estimate of the network performance, respectively. Also, in practice, the network may be well planed such that, at least, high-quality backhaul links are guaranteed. This is the motivation for our GA-based approach in which we propose a fairly simple network deployment optimization algorithm with no need for detailed mathematical analysis. This is important specially because
\begin{itemize}
    \item as we show in the following, with a well-planned network topology the need for routing, to compensate for temporal blockages, decreases which results in considerable implementation complexity reduction.
    \item Moreover, with our proposed GA-based approach it is possible to scale the network with proper deployment as more IAB nodes/\textcolor{black}{non-IAB backhaul link} connections are added to the network. 
    \item Finally, due to the generic characteristics of machine learning schemes, one can apply the same technique as our proposed GA method for both \textcolor{black}{non-IAB backhaul link} placement and SBS location optimization, as well as for the cases with different channel models/metrics of interest.
\end{itemize}

\textcolor{black}{It should be noted that, we are interested in the potential of optimal partial non-IAB type backhaul connections in order to find an upper bound on the performance of any real network that might be constrained. Such a constrained partial non-IAB type backhaul link deployment optimization would also be an interesting extension of this work, and any such network performance would be in between the optimized and the random partial non-IAB type backhaul link deployment.}

Particularly, in this paper, we propose two GA-based approaches \cite{genb1} to identify the optimal SBSs to be \textcolor{black}{non-IAB backhaul}-connected and the optimal locations for the SBSs, as explained in Algorithms 1 and 2, respectively. The algorithms are used to maximize the service coverage probability defined as the fraction of the UEs which have instantaneous UE data rates higher than or equal to a threshold $R_\text{th}$. That is, using \eqref{eq:14} and \eqref{eq:fibsbs}, the service coverage probability is given by
\begin{align}
 \rho= \Pr(R_\text{U} \ge \eta).
 \label{eq:rho}
\end{align}
\begin{figure}
\centerline{\includegraphics[width=3.5in]{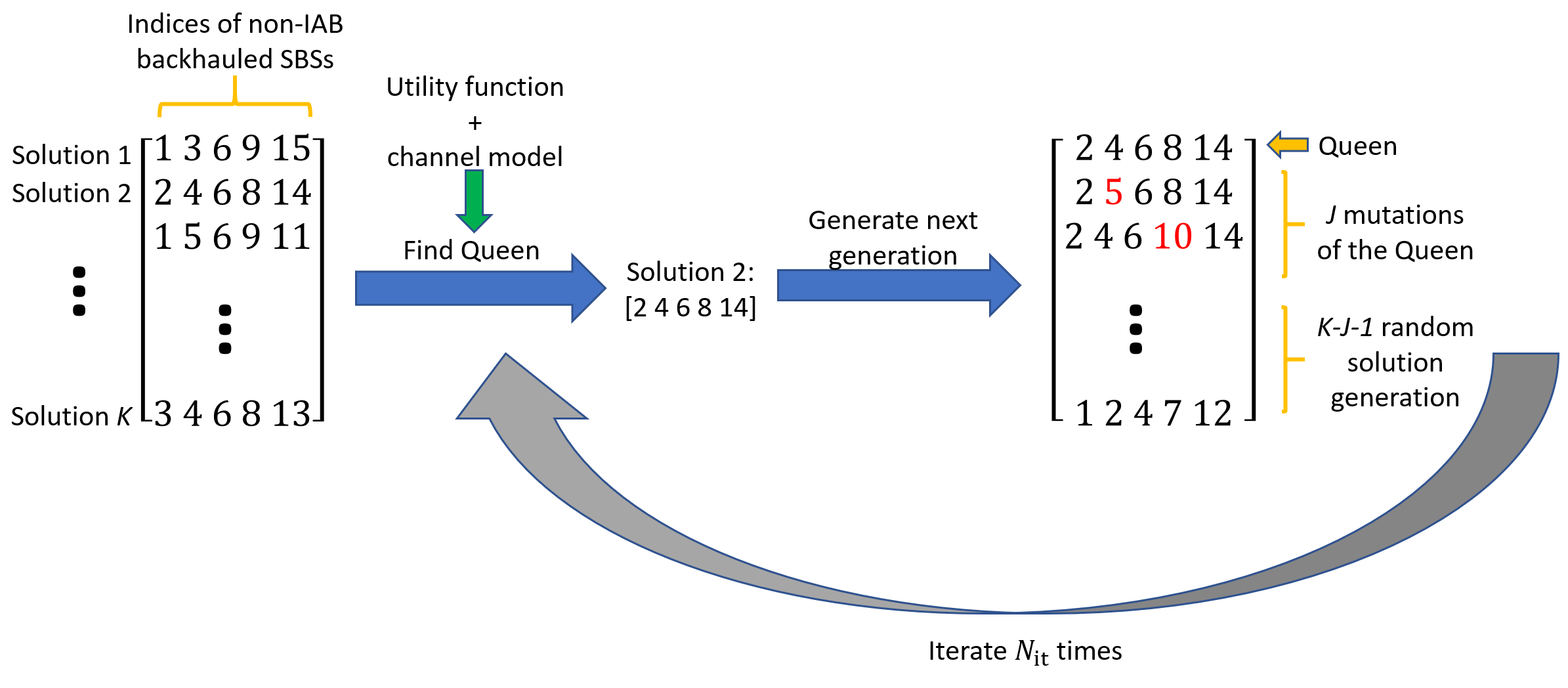}}
\caption{An example of the proposed GA in Algorithm 1 for \textcolor{black}{non-IAB backhaul link} distribution \textcolor{black}{between a fraction of} the SBSs. \textcolor{black}{In each iteration, the best solution (the Queen) is regenerated. Then, $J$ solutions are generated by small mutations in the Queen and $K-J-1$ possible solutions are generated randomly to avoid local minimums. The iterations continue for a number of times and the Queen of the last round is returned as the final solution.}
\label{iterationfig}}
\end{figure}
In words, both algorithms are based on the procedure described below. As shown in Fig. \ref{iterationfig}, we start the algorithm by considering $K$ possible selection strategies. For instance, Algorithm 1 considers $K$ possible SBS sets for \textcolor{black}{non-IAB backhaul link} placement and Algorithm 2 considers $K$ possible location sets for the SBSs. Then, in each iteration, we find the best strategy, i.e., selected solution, that maximizes the considered utility function, compared to the other $K-1$ selected strategies. This best strategy is referred to as the Queen. The Queen is considered as one of the possible solutions in the next iteration \textcolor{black}{of the algorithm} to guarantee the monotonic improvement of the algorithm performance in successive iterations. \textcolor{black}{That is, the Queen represents the regeneration operator in GA.} Also, for each iteration we create $J< K$ sets around the Queen. These matrices are created by applying slight modifications to the Queen, i.e., as a kind of mutation. For example, changing few SBSs of the set associated with the Queen generates these new sets needed for optimal SBS selection for \textcolor{black}{non-IAB backhaul link} placement in Algorithm 1. Also, in each iteration $K-J-1$ sets of selection strategies are generated randomly, to avoid the network to be trapped in a local minimum, and the iterations continue for $N_{\text{it}}$ iterations decided by the network designer \textcolor{black}{depending on the problem at hand (See Section \ref{res_sec})}. After running all considered iterations, the ultimate Queen is returned as the best selection rule for the current network instance. Particularly, Algorithm 1 returns the optimal SBS selection rule for \textcolor{black}{non-IAB backhaul link} placement, while Algorithm 2 returns the optimal SBS location selection rule. The suitable parameter setting for $K$, $J$ and $N_\text{it}$ in the algorithms can be obtained by the designer.

Considering Algorithms 1 and 2, the following points are interesting to note:
\begin{itemize}
    \item Our proposed algorithms result in significantly lower complexity, in comparison with the exhaustive search, as it only checks  $KN_{\text{it}}$ number of possible solutions \textcolor{black}{(see Section \ref{res_sec})}.
    \item  Moreover, due to Step 7 of the algorithms, where $K-J-1$ random possible solutions are checked in each iteration, the proposed algorithms mimic the exhaustive search if $N_\text{it} \to\infty$, and they reach the globally optimal selection rule if asymptotically many iterations are considered \textcolor{black}{ \cite{genb1}}.
    \item Unlike typical GAs, we do not use the crossover operation \textcolor{black}{and instead evaluate a few random solutions in each iteration.} This is because the proposed algorithms work well with no need for the additional complexity of the crossover operation, and converge with a few iterations (see Section \ref{res_sec}). However, it is straightforward to
include the crossover into the proposed algorithms where,
for instance, the Queen and the next best solutions are combined to generate new possible solutions.
    \item \textcolor{black}{The proposed algorithms optimize the network deployment off-line. However, it is straightforward to scale the network and adapt the algorithm in an on-line manner. For instance, adding new set of SBSs to an already-planned network deployment, one can rerun the algorithm for only a few iterations with the initial considered solutions not randomly but based on the Queen of the already-planned network. 
}
\end{itemize} 
\begin{algorithm}[tbph]
 \caption{GA-based \textcolor{black}{non-IAB Backhaul Link} Placement.}
 \begin{algorithmic}
\STATE \textit{In each network instance with a budget for} $N_{\text{f}}$ \textit{\textcolor{black}{non-IAB backhaul}-connected SBSs, and} $N_\text{s}>N_\text{f}$ \textit{SBSs, do the followings:} 
\begin{enumerate}[I.]
\item Consider $K$ sets of $N_{\text{f}}$ \textcolor{black}{non-IAB backhaul}-connected SBSs, $F_{k}$, and for each set create the corresponding channel matrix. Then, for each matrix $H_k$, $k=1...,K$, implement the system model in Section \ref{system_model}. 
  \item For each selected possible solution $F_k$, evaluate the objective function $U_k$, $k=1,...., K.$ For instance, considering the service coverage probability $\rho$ as the objective function, $U_k$ is given by \eqref{eq:rho}.
  \item Find the set of the SBSs among the considered solutions $F_k,\forall k,$ which gives in the best value of the objective function, service coverage probability (the Queen), e.g., $F_i$ where $\rho(H_k ) \leq \rho(H_i)$, $\forall k = 1, . . . ,K$. 
  \item $F_1 \longleftarrow F_i$
  \item Generate $J<K,$ sets of SBSs $F_j^\text{new}$, $j=1,...,J,$ around  the Queen, i.e., $F_i$. These sets of SBSs are generated by making small changes to the Queen, for instance, by replacing few SBSs with other SBSs.
  \item $F_{j+1} \longleftarrow F_j^\text{new}$, $j=1,...,J.$
  \item  Use the same procedure as in Step 1 and regenerate the remaining sets $F_j, j = J + 2,..., K$, randomly. 
  \item Proceed to Step 2 and continue the process for $N_{\text{it}}$ iterations pre-considered by the network designer.
  \\ \textit{Return the Queen as the optimal SBS selection rule for \textcolor{black}{non-IAB backhaul link} placement.}
  \end{enumerate}
 \end{algorithmic}
 \end{algorithm}

\textcolor{black}{Since the considered problem is polynomial time reducible, it is NP-hard \cite{nphard1}, \cite{nphard2}. Moreover, unless for the cases with very small networks, the search space increases rapidly with the network density which makes exhaustive search based optimization infeasible. As an example, the number of possible solution checkings of exhaustive search when optimizing the selection of non-IAB backhaul connected SBSs in a fixed network area is given by} 
\textcolor{black}{ 
\begin{equation}
S_{c}={\binom{N_\text{s}}{N_\text{f}}},
\end{equation}}
\textcolor{black}{
where ${n \choose k}$ denotes the "$n$ choose $k$" operator. In this way, for moderate/large values of $N_\text{f}$
and/or $N_\text{s}$, the search space soon becomes so large that exhaustive search is not feasible. However, the complexity of Algorithm 1 for a similar use case will be in the linear order of $KN_{\text{it}}$, reducing the complexity compared to exhaustive search significantly.}

Finally, it should be noted that: 
\begin{itemize}
    \item Depending on the infrastructures and the availability of \textcolor{black}{non-IAB backhaul link} connection, in practice it may not be possible to provide  some SBSs with a \textcolor{black}{non-IAB backhaul link} connection \textcolor{black}{(either fiber or a dedicated LoS nonIAB wireless backhaul).} This is because the \textcolor{black}{those} connections may be available in specific areas. In this way, \textcolor{black}{as explained in Section \ref{system_model}, }Algorithm 1 gives an optimistic ultimate network performance, as we consider no limitation for \textcolor{black}{non-IAB backhaul link} distribution among the SBSs. Then, depending on the specific network deployment, it is straightforward to adapt Algorithm 1 to consider restrictions on \textcolor{black}{non-IAB backhaul} link distribution among the SBSs.
    \item According to the 3GPP discussions, one can consider two different, namely, wide-area and local-area, IAB network deployments. Local-area IAB deployment refers to the cases with an unplanned network where the mobile terminal (MT) module of the IAB nodes have UE-type functionality, in terms of transmit power etc. Wide-area IAB network, on the other hand, refers to the cases with well-planned deployment and gNB-type functionalities for the IAB nodes. In this way, the proposed scheme mainly concentrates on the wide-area IAB network deployment, as the main use-case of the IAB networks.
\end{itemize}

 \begin{algorithm}[tbph]
 \caption{GA-based SBS Location Selection.}
 \begin{algorithmic}
 \STATE \textit{In each network instance with $N_{\text{s}}$ SBSs, from all possible locations in the space, do the followings:} 
 \begin{enumerate}[I.]
  \item Consider $K$ sets of $L_{k}$ locations, and for each set create the corresponding channel matrix $H_k$, $k=1...,K$, according to the system model in Section \ref{res_sec}. 
  \item Evaluate the objective function for each set, i.e., $U_k$, $k=1,...., K.$ For instance, considering the service coverage probability $\rho$ as the objective function, $U_k$ is given by \eqref{eq:rho}.
  \item Find the Queen, i.e., the set of locations which gives the best value of the objective function, i.e., service coverage probability, among the considered sets, e.g., $L_i$ where $\rho(H_k ) \leq \rho(H_i)$, $\forall k = 1, . . . ,K$, 
  \item $L_1 \longleftarrow L_i$
  \item Generate $J<K,$ sets of locations $L_j^\text{new}$, $j=1,...,J$, around $L_i$. These sets of locations are generated by making small changes to the Queen, for instance, by replacing few locations with another sets of locations.
  \item $L_{j+1} \longleftarrow L_j^\text{new}$, $j=1,...,J.$
  \item  Use the same procedure as in Step 1 and regenerate the remaining sets $L_j, j = J + 2,..., K$, randomly. 
  \item Proceed to Step 2 and continue the process for $N_{\text{it}}$ iterations pre-considered by the network designer.
  \\ \textit{Return the Queen as the optimal SBS location selection rule.}
   \end{enumerate}
 \end{algorithmic}
 \end{algorithm}
 
\section{Performance Evaluation Of Deployment Optimization}
\label{res_sec}
The simulation results and discussions are divided into three main areas in which 1) we evaluate the convergence behaviour of the proposed algorithms, and we study their effect on optimizing the IAB network performance, 2) verify the effect of environmental parameters on the coverage probability, and 3) evaluate the system performance for different transmission capabilities of the nodes. Then, in Section \ref{routing_sec}, we investigate the effect of routing on the performance of IAB networks experiencing temporal blockings.

The general system parameters are presented in Table \ref{tab1} and, in each figure, we give the detailed system parameters in the figure captions. The IAB network is deployed in a 2D disk, in which the blockage, and the tree distributions are also modelled using statistical models described in Section \ref{system_model}. In particular, the network is a hybrid IAB deployment, of which a fraction of the SBSs will be \textcolor{black}{non-IAB backhaul}-connected while the others are backhauled using IAB.  \textcolor{black}{In all figures, except for Fig. \ref{treefol} which studies the system performance in \textcolor{black}{suburban} areas, we focus on dense areas as the most important use-case in IAB networks. \textcolor{black}{Also, in all figures, except in Fig. \ref{gainpl}, we ignore interference in the backhaul links, and assume them to be noise-limited.} In Figs. \ref{fibiter},\ref{blockingpl}, \ref{treefol}-\ref{cdfpl}, we study the system performance in the cases with \textcolor{black}{non-IAB backhaul link} placement optimization (Algorithm 1). \textcolor{black}{Figures \ref{lociter}, \ref{powerpl}, \ref{cdfpl_loc} and \ref{temp_covg} present the results for the cases with Algorithm 2 optimizing the SBSs locations.}}
\begin{table}
\caption{Simulation Parameters.}
\setlength{\tabcolsep}{3pt}
\begin{tabular}{|p{95pt}|p{150pt}|}
\hline
Parameters& 
Value \\
\hline
 Carrier frequency&28 GHz \\Bandwidth& 1 GHz\\ IAB node and UEs density& \{MBS, SBS, UE\} = (2, 50, 500) /$\text{km}^2$\\ Blocking density&  500 /$\text{km}^2$\\Path loss exponents& \{LoS, NLoS\} = (3, 4)\\ Main lobe antenna gains& \{MBS, SBS, UE\} = (18, 18, 0) dBi\\ Side lobe antenna gains& \{MBS, SBS, UE\} = (-2, -2, 0) dBi\\ Half power beamwidth&  {$\text{30}$}\\ Noise power& 5 dB\\ Percentage of \textcolor{black}{non-IAB backhauled} SBS nodes& 10\%\\ In-leaf percentage& 15\%\\ Tree depth& 7.5 m

\\

\hline

\end{tabular}
\label{tab1}
\end{table}

\subsection{On the Performance of the Proposed Algorithms}
\begin{figure}
\centerline{\includegraphics[width=3.5in]{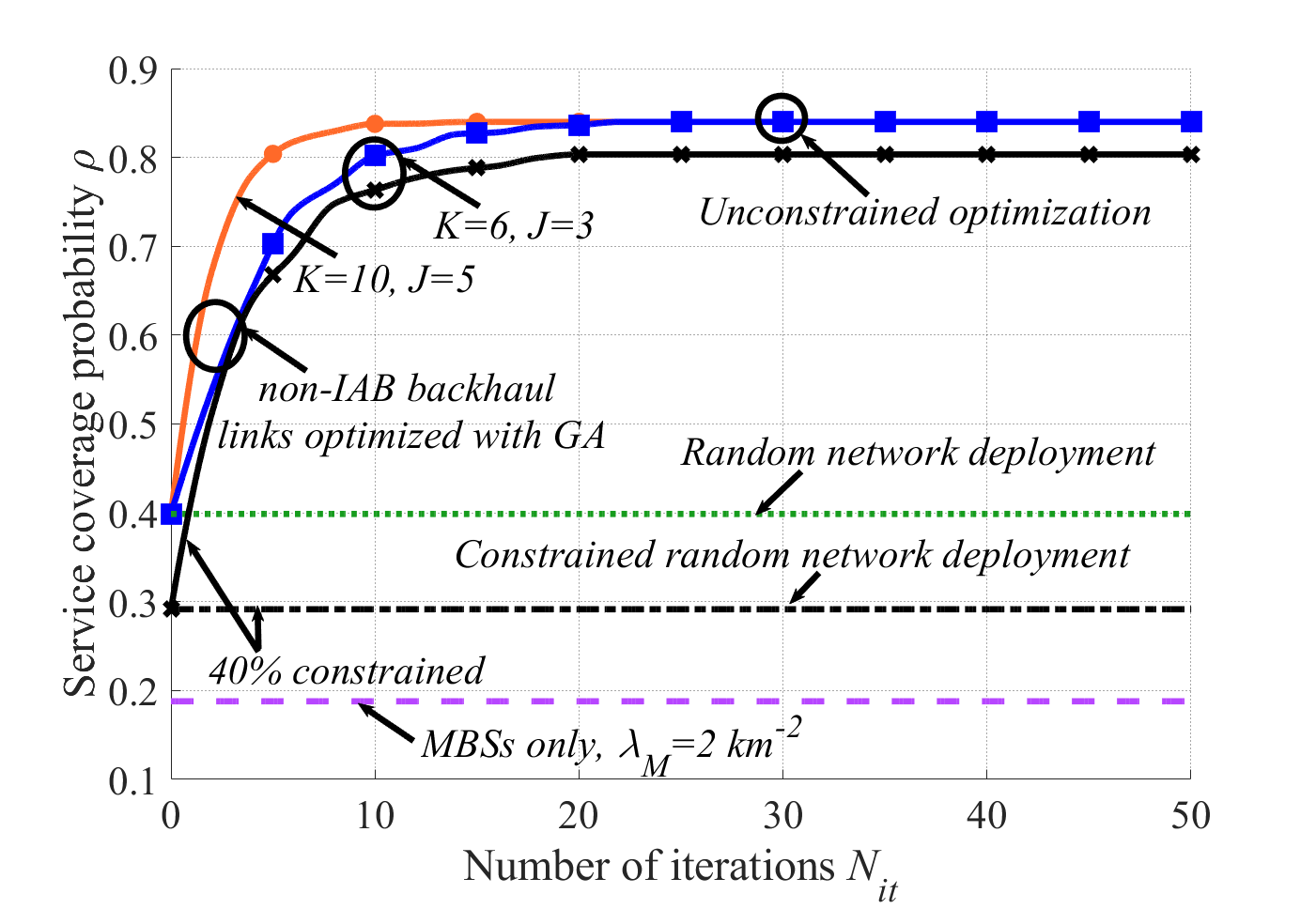}}
\caption{Service coverage probability as a function of the number of iterations in Algorithm 1 with non-IAB backhaul link connection distribution, and ${P_{\text{m}}, P_{\text{s}}, P_{\text{u}}} = (40, 24, 0)$ dBm. The parameters are set to $\lambda_\text{M}=2$ $\text{km}^{-2}$, $\lambda_\text{S}=50$ $\text{km}^{-2}$ and $\lambda_\text{U}=500$ $\text{km}^{-2}$. \textcolor{black}{The results are presented in both cases with constrained and free non-IAB backhaul link distribution in the coverage area.}\label{fibiter}}
\end{figure}

\begin{figure}
\centerline{\includegraphics[width=3.5in]{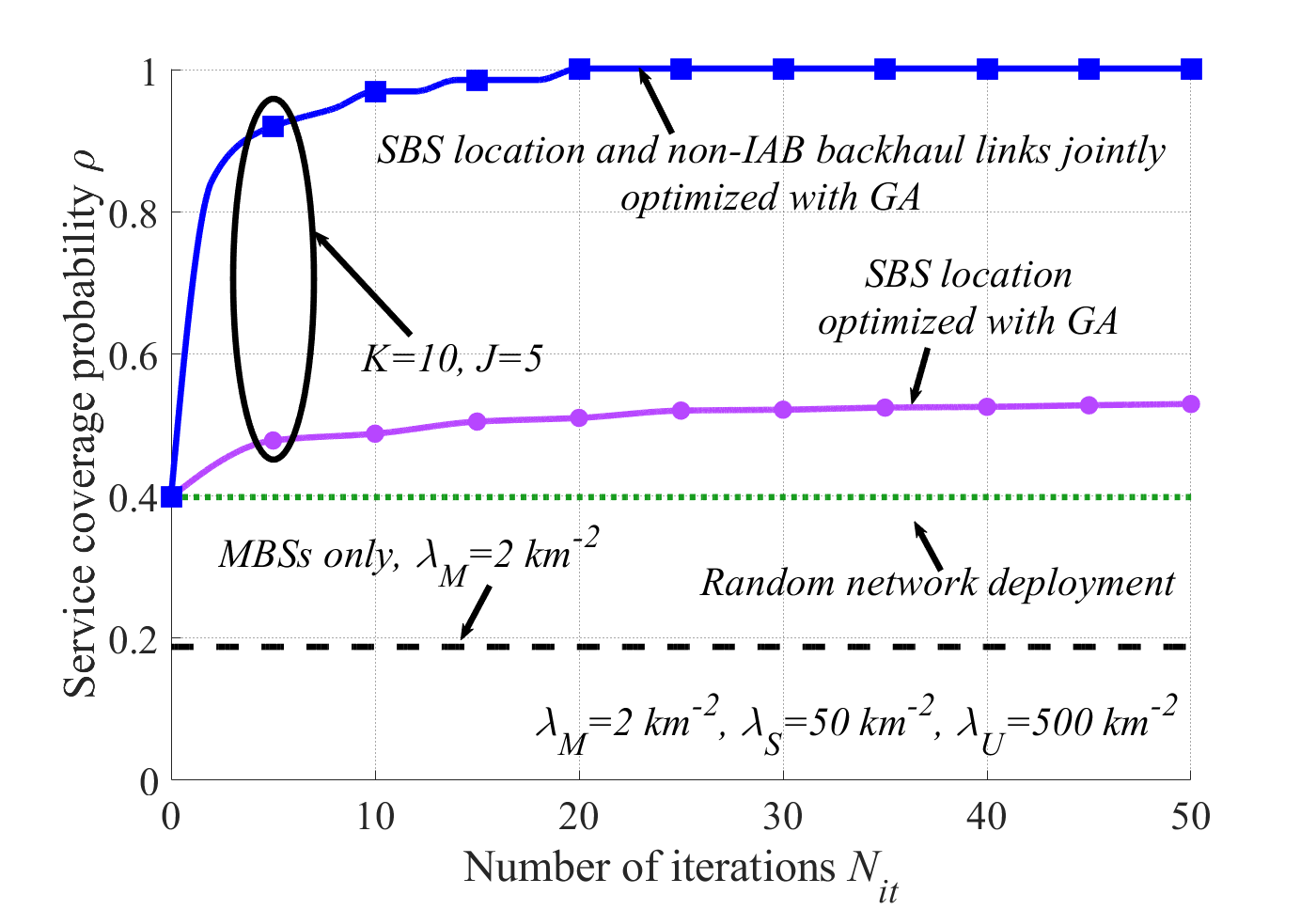}}
\caption{\textcolor{black}{Service coverage probability as a function of the number of iterations in Algorithm 2 with IAB node placement optimization, IAB node placement and non-IAB backhaul links joint optimization and ${P_{\text{m}}, P_{\text{s}}, P_{\text{u}}} = (40, 24, 0)$ dBm. The parameters are set to $\lambda_\text{M}=2$ $\text{km}^{-2}$, $\lambda_\text{S}=50$ $\text{km}^{-2}$ and $\lambda_\text{U}=500$ $\text{km}^{-2}$.} \label{lociter}}
\end{figure}

In Figs. \ref{fibiter}-\ref{lociter}, we study the convergence performance of the proposed algorithms, and compare the results with the cases having only MBSs or random network deployment.
Figure. \ref{fibiter} shows the service coverage probability achieved for different numbers of iterations in Algorithm 1 with optimal \textcolor{black}{non-IAB backhaul link} connection distribution and different algorithm parameters $K$ and $J$. Here, the results are presented for the cases with 10\% of the SBSs having the possibility to be \textcolor{black}{non-IAB backhaul}-connected. Then, Fig. \ref{lociter} demonstrates the IAB network service coverage probability as a function of the number of iterations in Algorithm 2, and compares the results with the benchmark schemes using only MBSs or random network deployment of which 10\% of the SBSs are \textcolor{black}{non-IAB backhaul}-connected.

\textcolor{black}{As in every machine learning-based algorithm applied in large systems, the main challenge of the proposed scheme is to achieve reasonably good results within limited iterations. This is challenging specially in the cases with large network density and/or joint optimization of the non-IAB backhaul links distribution and IAB nodes placements, as the search space increases rapidly. However, as shown in the evaluations (Figs. \ref{fibiter}-\ref{lociter}), with a proper setting of the algorithms parameters, the GA converges with a few number of iterations.}

As seen in Figs. \ref{fibiter} and \ref{lociter}, the developed Algorithms 1 and 2 converge rapidly to give a maximum service coverage probability. For example, Fig. \ref{fibiter} converges with almost $N_\text{it}=20$ iterations which, with $K=6$, leads to a total of 120 possible solution checkings. \textcolor{black}{As a result, the proposed algorithm reduces the complexity compared to exhaustive search significantly because with $\lambda_\text{s}=50 \text{ km}^{-2}$ and the network area of $1 \text{ km}^{-2}$ exhaustive search requires ${\binom{50}{5}}\simeq 2\times 10^6$ solution checkings, i.e. $\simeq 17000$ times larger search than those in our proposed scheme.} In particular, the proposed algorithms have improved the service coverage probability, compared to the IAB network with random node locations and random non-IAB backhaul connections, significantly. For instance, with the parameter settings of Fig. \ref{fibiter}, optimizing the non-IAB backhaul link distribution among $10\%$ of the SBSs increases the coverage probability from 40\% with random non-IAB backhaul link distribution to 85\%. \textcolor{black}{Moreover, with the parameter settings of Fig. \ref{lociter}, optimizing the SBSs location leads to a coverage probability increment from 40\% with random network deployment to 55\%, while the joint optimization of non-IAB backhaul link distribution and SBS location further improves the coverage probability reaching the maximum of 100\%.}

\textcolor{black}{In the simulations, we considered no constraints on the SBSs and non-IAB backhaul links locations. However, in practice, it may not be possible to place the SBSs and the non-IAB backhaul links freely. To evaluate this point, in Fig. \ref{fibiter} we study the system performance in the cases with constraints on \textcolor{black}{the non-IAB backhaul link distribution}. Particularly, Fig. \ref{fibiter} shows the coverage probability in the cases where the non-IAB backhaul links can not be placed in 40\% of the area selected randomly. Here, the results presented for both cases with random and optimized distributions of the non-IAB backhaul links in the 60\% of the coverage area. As seen in Fig. 8, although the constraint on topology optimization may limit the benefit of IAB, still the system performance is improved compared to the cases with only MBSs. Also, for a broad range of parameter settings, the effect of topology constraints on the network performance is not significant.}

Finally, as expected and also demonstrated in Figs. \ref{fibiter}-\ref{lociter}, as the UEs density increases, MBSs alone can not support the UEs’ coverage probability requirements, and indeed we need to densify the
network using (IAB) nodes of different types. In this way, as also experienced in practical network implementations, a well-planned network deployment results in significant performance improvement, which reduces the need for high network node density as well as the implementation cost.

Note that, while Figs. \ref{fibiter}-\ref{lociter} show monotonic improvement of the system performance in successive iterations, in some iterations the proposed algorithms may follow a ladder-shape convergence pattern. This is because the service coverage probability does not necessarily improve in each iteration, and there is a possibility to reach a local optimum in some iterations. However, we always elude the local minima due to Step 7 of Algorithms 1 and 2. Thus, given that sufficiently large number of iterations are carried out, the algorithm converges to a (sub)optimal solution.  

\begin{figure}
\centerline{\includegraphics[width=3.5in]{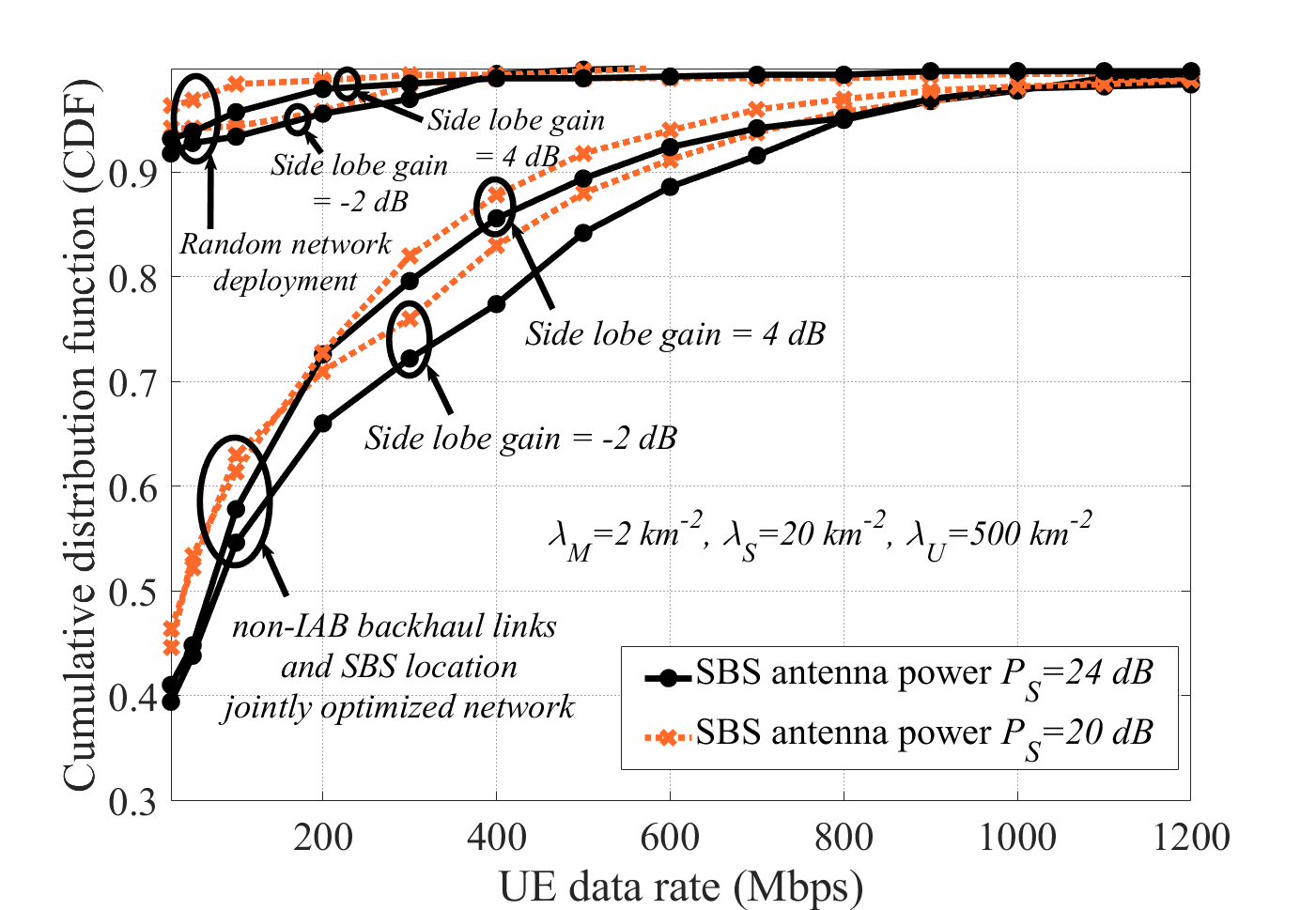}}
\caption{\textcolor{black}{CDF of the achievable rates with  $P_{\text{m}}, P_{\text{u}} = (40, 0)$ dBm for SBS location and non-IAB backhaul links joint optimization. The parameters are set to $\lambda_\text{M}=2$ $\text{km}^{-2}$,  $\lambda_\text{S}=20$ $\text{km}^{-2}$ and $\lambda_\text{U}=500$ $\text{km}^{-2}$.}}
\label{cdfpl}
\end{figure}

\begin{figure}
\centerline{\includegraphics[width=3.5in]{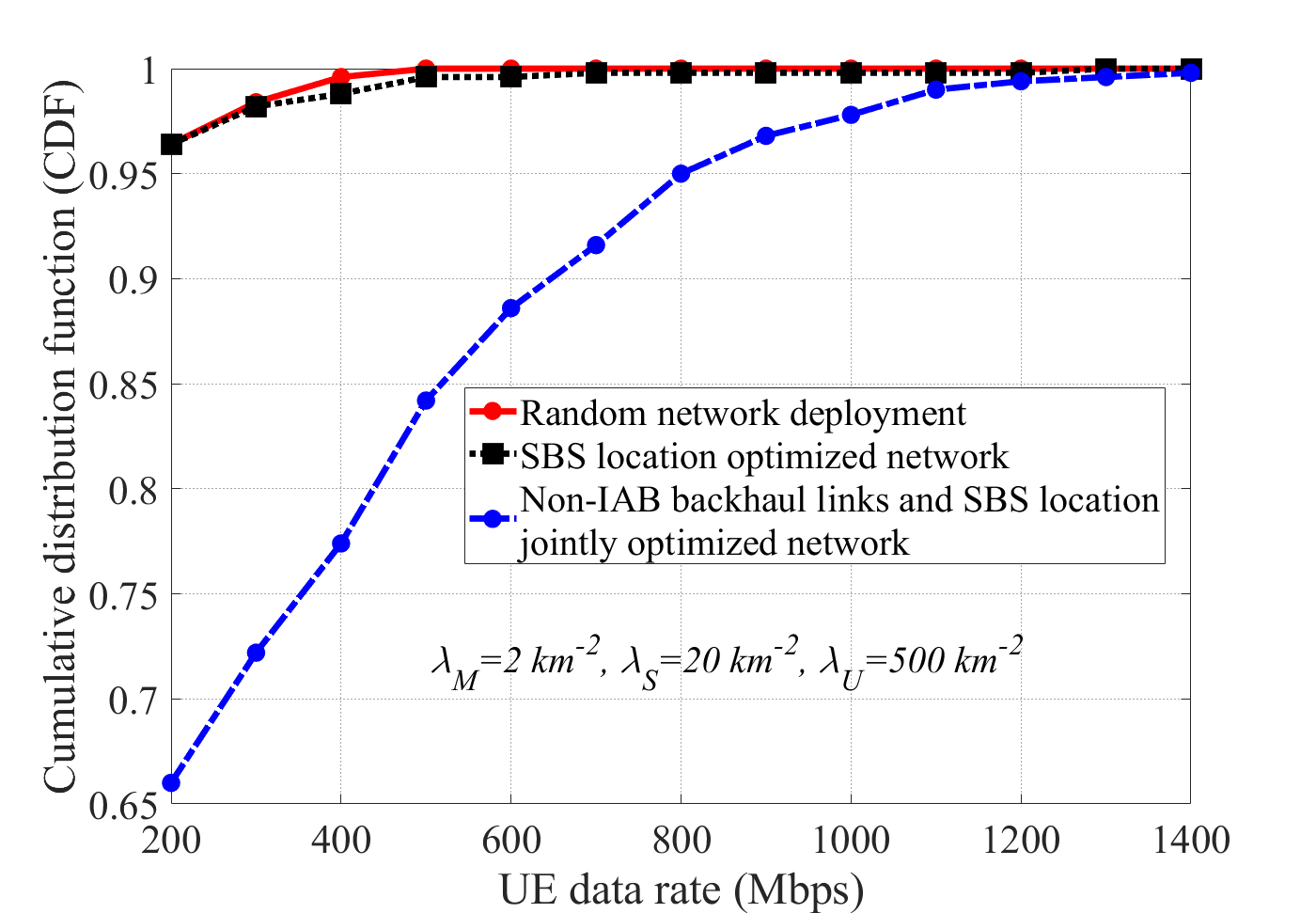}}
\caption{\textcolor{black}{CDF of the achievable rates with  $P_{\text{m}}, P_{\text{u}} = (40, 0)$ dBm for SBS location optimization, SBS location and non-IAB backhaul links joint optimization. The parameters are set to $\lambda_\text{M}=2$ $\text{km}^{-2}$,  $\lambda_\text{S}=20$ $\text{km}^{-2}$ and $\lambda_\text{U}=500$ $\text{km}^{-2}$.}}
\label{cdfpl_loc}
\end{figure}

\textcolor{black}{
In Figs. \ref{cdfpl} and \ref{cdfpl_loc}, we study the cumulative distribution function (CDF) of the UEs achievable data rates in the cases with SBS location optimization as well as joint non-IAB backhaul link distribution and SBS location optimization, and compare the results with random network deployment. Here, the parameters are set to $\lambda_\text{M}=2$ $\text{km}^{-2}$,  $\lambda_\text{S}=20$ $\text{km}^{-2}$ and $\lambda_\text{U}=500$ $\text{km}^{-2}$, and in all cases $10\%$ of the SBSs are non-IAB backhaul-connected. Also, the joint optimization follows the same setup as in Algorithms 1-2.}

\textcolor{black}{
 \textcolor{black}{As can be seen in Fig. \ref{cdfpl}, with a random deployment and the parameter settings of the figure, (almost) all UEs maximum achievable rates are below $400$ Mbps, the result which holds for both considered values of the IAB nodes transmit powers and side lobe gains.} On the other hand, jointly optimizing the non-IAB backhaul link distribution and SBSs locations gives the chance to support higher access data rates, depending on the UEs position and their associated backhaul links qualities. For instance, as opposed to the cases with random network deployment, with $P_\text{s}=24$ dBm around $25\%$ of the UEs may experience $>400$ Mbps access rates, if the non-IAB backhaul \textcolor{black}{link distribution and the SBSs locations}  are properly planned (Fig. \ref{cdfpl}).}

\textcolor{black}{In our simulations, we consider relatively low side lobe gains, compared to the main lobe gain. This is motivated by the fact that, to guarantee high-rate reliable backhaul performance at mmw spectrum, IAB nodes are expected to be equipped with a large number of antennas and be capable of directional beamforming. However, depending on the hardware properties, in practice there may be cases with relatively high side lobe gains \cite{steinmetzer2017compressive}. For this reason, in Fig. \ref{cdfpl} we study the effect of the side lobe gain on the network performance, and verify the coverage probability for different values of side lobe gains. As demonstrated in the figure, while the coverage probability is slightly reduced by increasing the side lobe gain, for a broad range of parameter settings, the relative performance loss is negligible.}

\textcolor{black}{
In harmony with Fig. \ref{cdfpl}, Fig. \ref{cdfpl_loc} shows that, the SBS location-optimized network can support UEs data rates up to 1200 Mbps while non-IAB backhaul link distribution and SBS locations jointly optimized network can support UEs data rates up to around 1400 Mbps. In this way, the joint optimization improves
the performance, compared to optimizing one of the parameters, at the  cost of higher computational
complexity. Finally, note that, along with simplifying the optimization process, one of
the motivations for separate optimization of the SBSs locations and non-IAB backhaul
links distribution is that in practice the SBSs and the non-IAB backhaul links many be
deployed by different companies or in different times, which makes joint optimization
difficult.}

\subsection{Effect of Blocking and Tree Foliage}
\begin{figure}
\centerline{\includegraphics[width=3.5in]{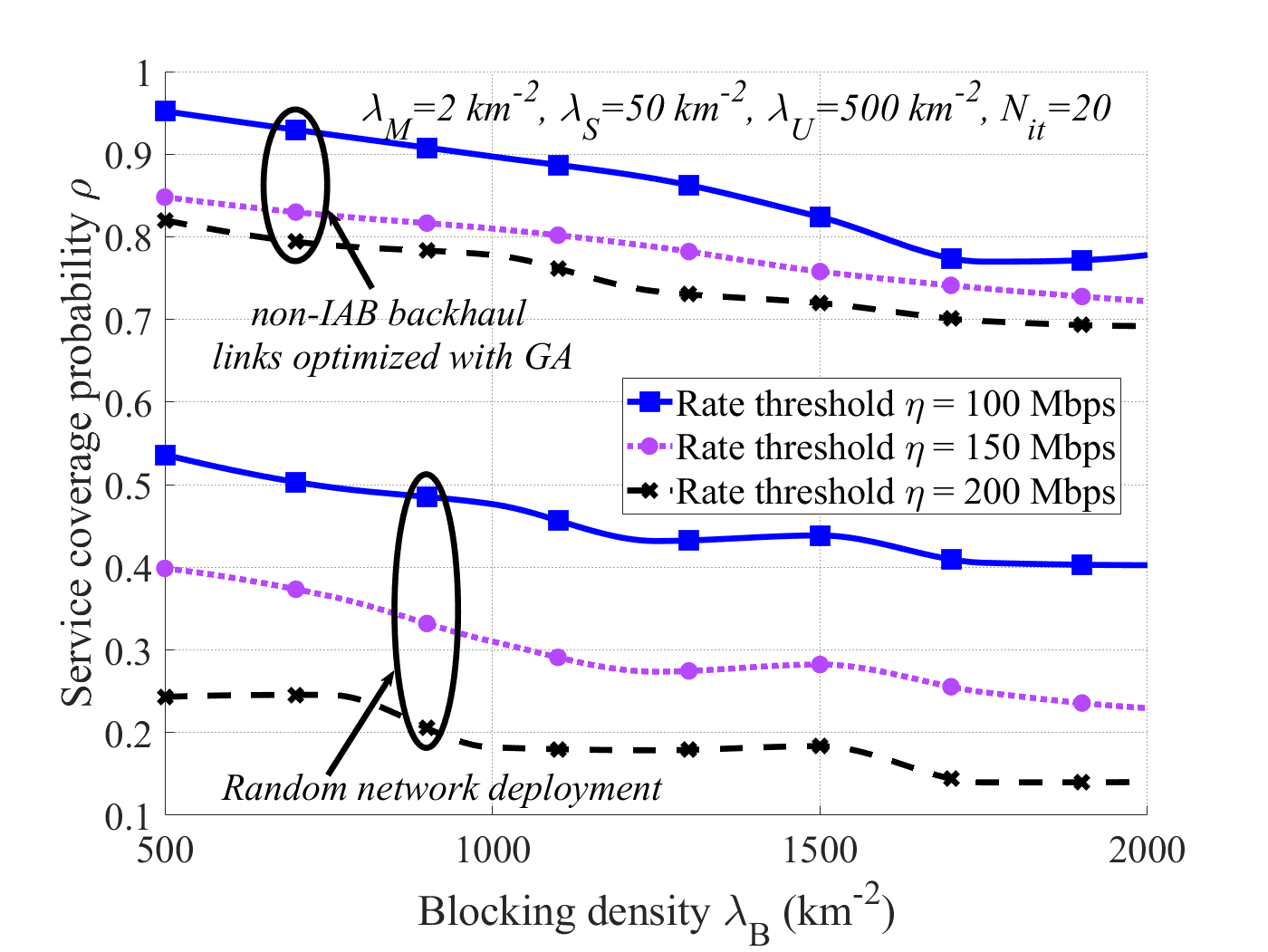}}
\caption{Service coverage probability of the IAB network as a function of the blocking density $\lambda_\text{B}$, with ${P_{\text{m}}, P_{\text{s}}, P_{\text{u}}} = (40, 24, 0)$ dBm and different methods of non-IAB backhaul link connection distribution among $10\%$ of the SBSs. The parameters are set to $\lambda_\text{M}=2$ $\text{km}^{-2}$, $\lambda_\text{S}=50$ $\text{km}^{-2}$, $\lambda_\text{U}=500$ $\text{km}^{-2}$ and $N_\text{it}=20$.  \label{blockingpl}}
\end{figure}
\begin{figure}
\centerline{\includegraphics[width=3.5in]{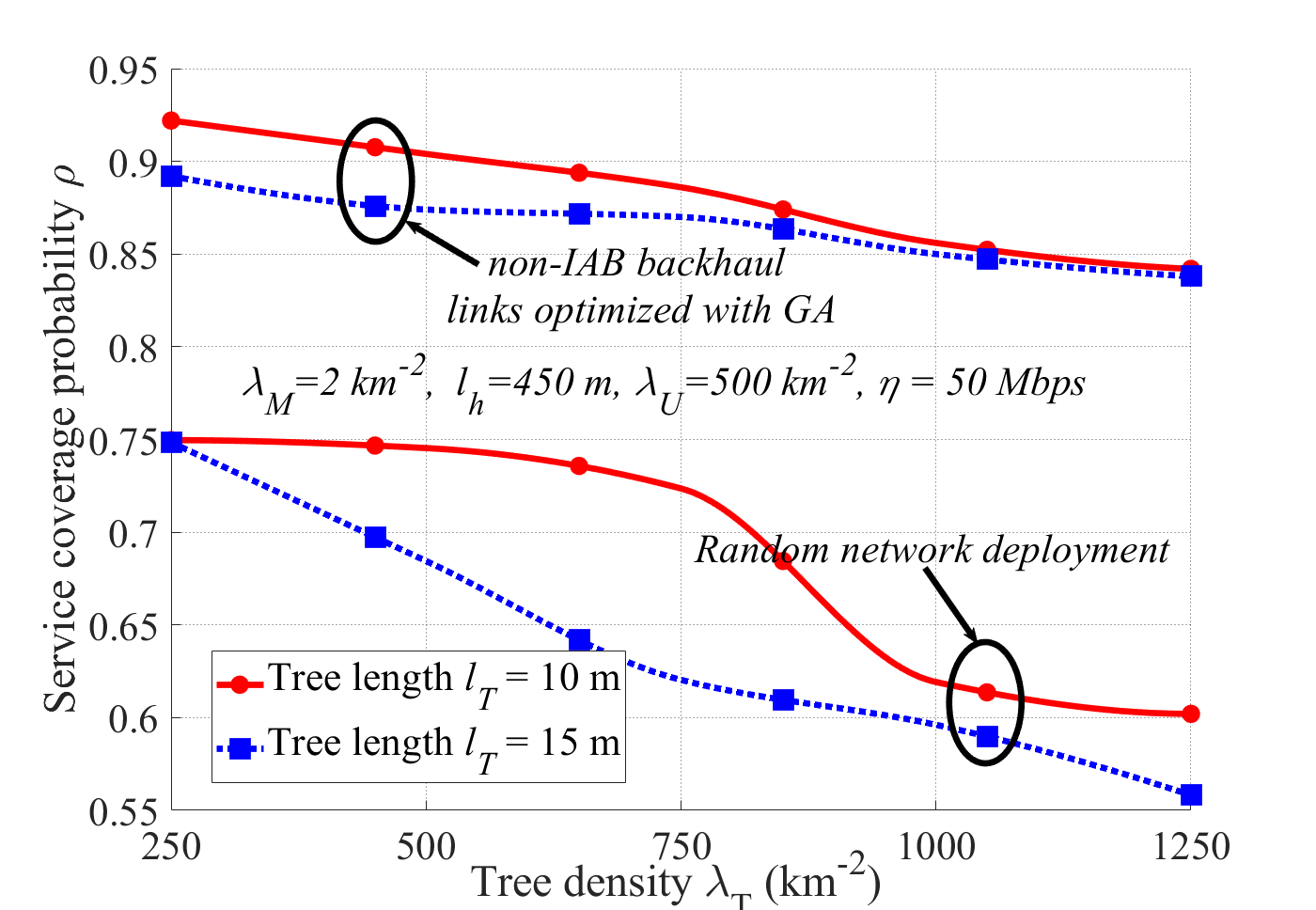}}
\caption{Service coverage probability of the IAB network as a function of tree density $\lambda_\text{T},$ with  $P_{\text{m}}, P_{\text{s}}, P_{\text{u}} = (40, 33, 0)$ dBm and different methods of \textcolor{black}{non-IAB backhaul link} connection distribution among $10\%$ of the SBSs. The parameters are set to $\lambda_\text{M}=2$ $\text{km}^{-2}$, $l_\text{h}=450$ $\text{m}$, $\lambda_\text{U}=500$ $\text{km}^{-2}$ and $\eta=50$ $\text{Mbps}$.  
\label{treefol}}
\end{figure}
In contrast to the \textcolor{black}{non-IAB backhaul}-connected networks, IAB networks may be affected by environmental effects specially the blockage and the tree foliage\footnote{As reported in [25], with the typical hop lengths of the IAB networks and 28 GHz, the effect of the rain on the coverage probability of IAB network is negligible.}. In Figs. \ref{blockingpl} and \ref{treefol}, we respectively study the effect of the blockage and tree foliage on the coverage probability of the IAB network with random deployment or GA-optimized \textcolor{black}{non-IAB backhaul link} distribution. Here, the results are presented for different rate thresholds of the UEs, i.e., $\eta$ in (15). In particular, Fig. \ref{blockingpl} shows the service coverage probability considering the FHPPP-based germ-grain blockage model for different blocking densities.

Although urban areas are the main point of interest for IAB network, to study the potentials of its usage in suburban areas, in Fig. \ref{treefol} we demonstrate the service coverage probability as a function of the tree density in the suburban areas. Here, we present the results for the average hop distance $l_\text{h}$ = 450 m corresponding to SBSs density, $\lambda_\text{S}$ = 8 $\text{km}^{-2}$. This is motivated by, e.g., \cite{erefaddnewc}, reporting the tree foliage as one of the main challenges of IAB in suburban areas. According to Figs. \ref{blockingpl}-\ref{treefol}, the following points can be concluded:

\begin{itemize}
   
\item The GA-based planned deployment shows significant improvement and resilience to blockage and tree foliage, compared to random deployment, where the coverage probability is not \textcolor{black}{much} affected by the blockage (Fig. \ref{blockingpl}). For instance, with the given system configuration in Fig. \ref{blockingpl}, the GA optimized setup shows 0.85 service coverage probability at $\eta =150$ Mbps, $\lambda_{\text{B}}$ = 1000 $\text{km}^{-2}$, while the coverage probability reduces to 0.72 at $\lambda_{\text{B}}$ = 2000 $\text{km}^{-2}$, i.e., only 15\% coverage loss by doubling the blockage density. On the other hand, with a random network deployment, $\eta = 150$ Mbps and $\lambda_\text{B}=1000$, the coverage probability is only 0.4 and it is dropped to 0.23, i.e., 42\% performance degradation, as the blockage density increases to $\lambda_\text{B}$ = 2000 $\text{km}^{-2}$.

\item In suburban area and with a random network deployment, the coverage probability is considerably affected by the tree foliage, especially when the trees density and/or length increase. However, we note that the introduction of GA optimization on selecting the SBSs with \textcolor{black}{non-IAB backhaul}-connection has brought resilience to the tree foliage. This is due to the fact that the algorithm finds the optimum set of nodes minimizing the SBS links with high losses due to tree foliage. For instance, considering the settings of Fig. \ref{treefol} and the random FHPPP model with $l_T$ = 15 m, the service coverage probability drops from 0.75 to 0.55 (26\% coverage degradation) when the tree density is increased from 250 to 1250 $\text{km}^{-2}$. However, the same tree density increase at $l_T$ = 15 m in GA-optimized network gives a drop only from 0.88 till 0.83, i.e., only $5\%$ performance drop, the result which is almost independent of the tree length.

\end{itemize}

 In general, the robustness of IAB in the presence of tree foliage is hard to predict due to the fact that the link quality can vary depending on the characteristics of the tree lines. Particularly, the backhaul links quality may change due to wet trees, snow on the trees, wind and varying percentage of leaves in different seasons. However, we conclude that, although the IAB is prone to medium/highly densified tree foliage in suburban areas, network planning can reduce much of its adverse effect, and the mmWave IAB is expected to work well for areas with low/moderate foliage level.
 
\subsection{Effect of Antenna Gain and Transmit Power}
\begin{figure}
\centerline{\includegraphics[width=3.5in]{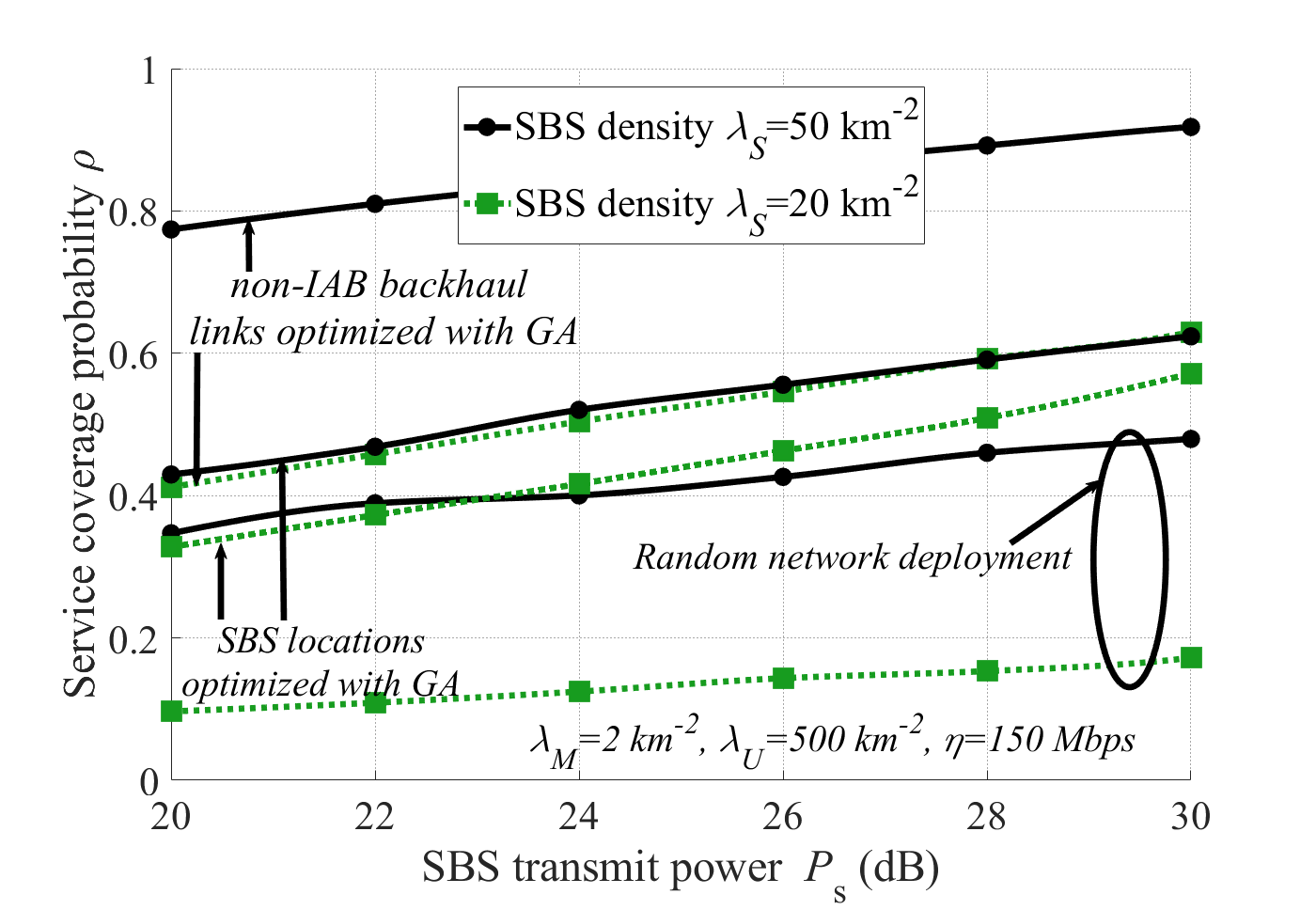}}
\caption{Service coverage probability of the IAB network as a function of the SBSs transmit power $P_{\text{s}}$, with ${P_{\text{m}}, P_{\text{u}} = (40, 0)}$ dBm  for both \textcolor{black}{non-IAB backhaul link} location and node placement optimization methods, $\lambda_\text{M}=2$ $\text{km}^{-2}$, $\lambda_\text{U}=500$ $\text{km}^{-2}$ and $\eta=150$ $\text{Mbps}$.}
\label{powerpl}
\end{figure}

\begin{figure}
\centerline{\includegraphics[width=3.5in]{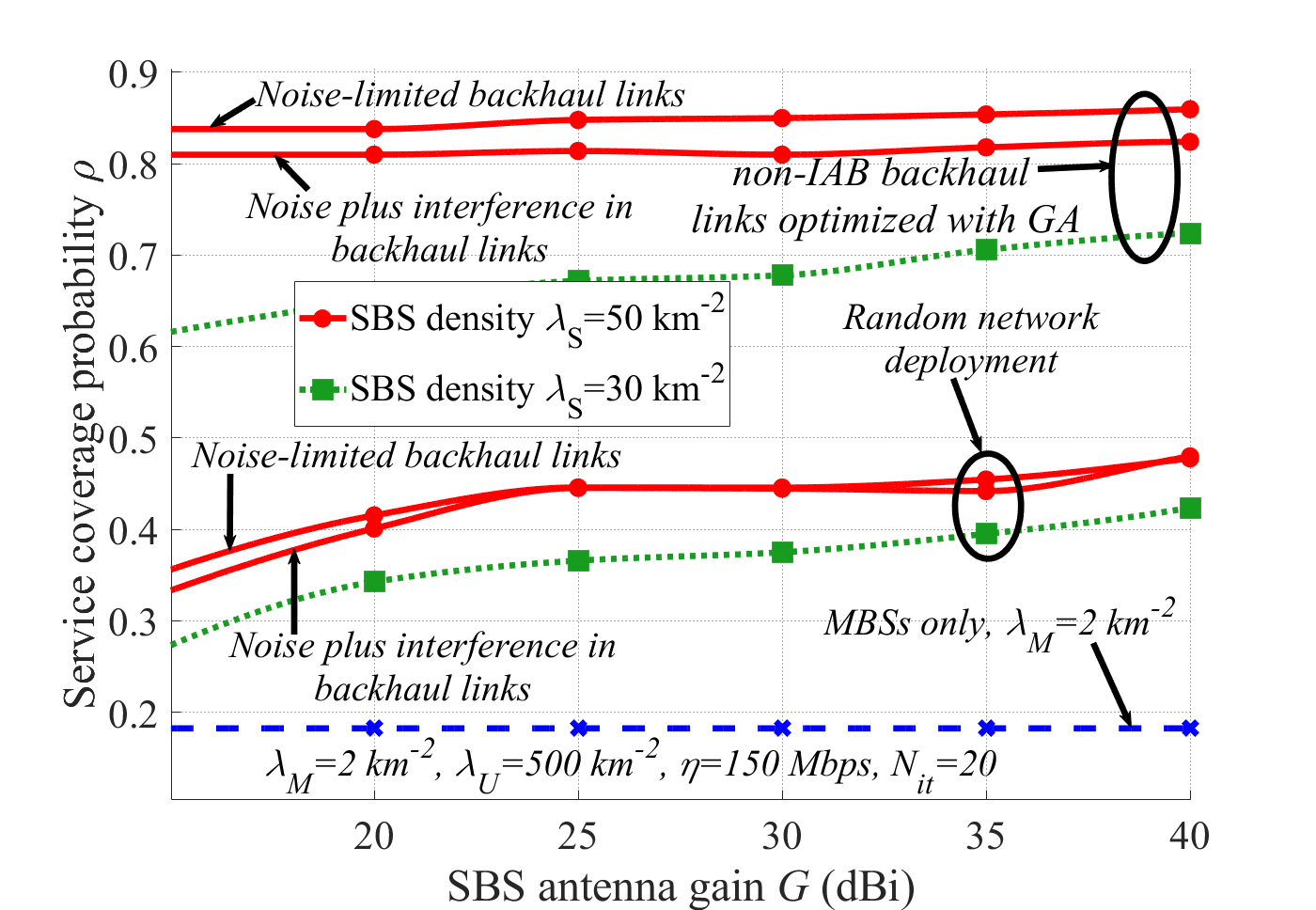}}
\caption{Service coverage probability of the IAB network as a function of the SBSs antenna gain $G$, with ${P_{\text{m}}, P_{\text{s}}, P_{\text{u}} = (40, 24, 0)}$ dBm  for \textcolor{black}{non-IAB backhaul link} location optimization, $\lambda_\text{M}=2$ $\text{km}^{-2}$, $\lambda_\text{U}=500$ $\text{km}^{-2}$, $\eta=150$ $\text{Mbps}$ and $N_\text{it}=20$. \label{gainpl}}
\end{figure}

In Fig. \ref{powerpl}, we demonstrate the service coverage probability as a function of the SBS transmit power for three scenarios, namely, random FHPPP-based deployment, GA-based \textcolor{black}{non-IAB backhaul link} distribution and GA-based SBS location optimization. Also, Fig. \ref{gainpl} shows the service coverage probability as a function of the SBS antenna gain for random FHPPP-based deployment with $10\%$ \textcolor{black}{non-IAB backhaul}-connected SBSs, macro-only network and GA-optimized \textcolor{black}{non-IAB backhaul link} distribution between $10\%$ of the SBSs. \textcolor{black}{In addition, to verify the effect of interference in the backhaul links, the figure shows the service coverage probability in the presence of both noise-limited and noise plus interference limited backhaul links.} Here, we increase the SBS antennas' main lobe gain, while fixing the side lobe gain at -2 dB. 

As we observe in Fig. \ref{powerpl}, both GA-optimization methods used for selecting the dedicated \textcolor{black}{non-IAB backhauled} nodes and selecting SBSs locations have significantly increased the system coverage probability, compared to random deployment, and the relative effect of network planning increases with the SBSs' transmit power (Fig. \ref{powerpl}). Moreover, with different deployment conditions and the considered range of transmit powers, the coverage probability increases almost linearly with the SBSs transmit power, while the relative benefit of the transmit power increment increases in the cases with a well-planned network (Fig. \ref{powerpl}). Also, Fig. \ref{gainpl} demonstrates that, for the considered parameter setting of the figure and moderate/high antenna gains, the system performance is almost insensitive to the antenna gain specially if the network is well planned. \textcolor{black}{Finally, as seen in Fig. \ref{gainpl}, the impact of the interference in the backhaul links is negligible, and thus, the backhaul links can be well assumed to be noise-limited} (Also, see \cite{eref17,eref18,eref19}, \cite{erefour} for further discussions).

\begin{figure}
\centerline{\includegraphics[width=3.5in]{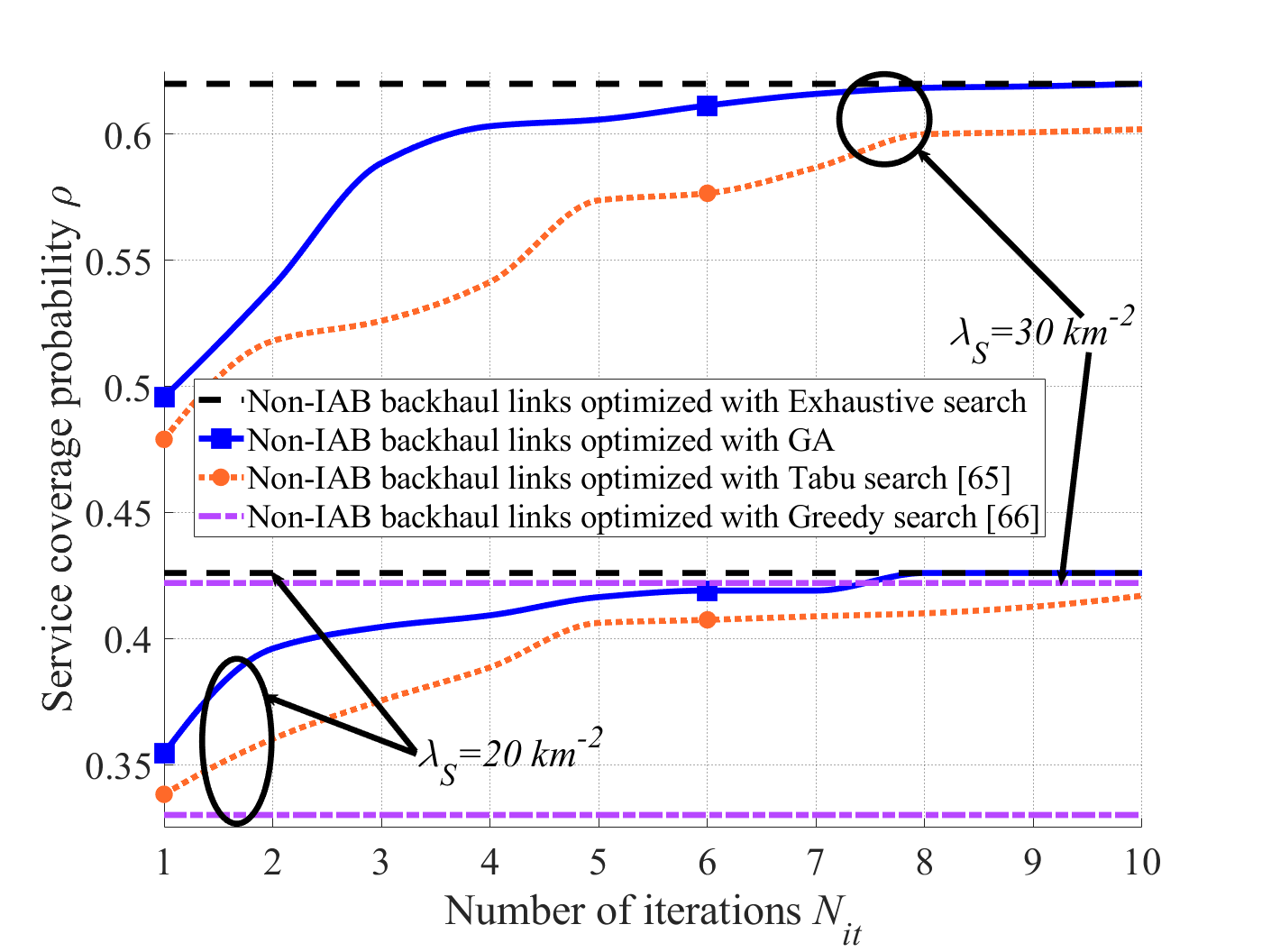}}
\caption{\textcolor{black}{Service coverage probability as a function of the number of iterations in Algorithm 1 with non-IAB backhaul links optimization. The parameters are set to ${P_{\text{m}}, P_{\text{s}}, P_{\text{u}}} = (40, 24, 0)$ dBm, $\lambda_\text{M}=2$ $\text{km}^{-2}$, and $\lambda_\text{U}=500$ $\text{km}^{-2}$.}}
\label{alg_heuristic}
\end{figure}

\textcolor{black}{
In Fig. \ref{alg_heuristic}, we compare the coverage probability of the proposed GA-based scheme with those obtained by different state-of-the-art algorithms including exhaustive search, Tabu algorithm \cite{z_1} and Greedy algorithm \cite{z_3}.  Note that Tabu is an evolutionary algorithm with a specific method of producing the generations (see \cite{z_2} for details) while with Greedy algorithm, e.g., the non-IAB backhaul links locations are determined one-by-one \cite{z_3}. Here, the parameters are set to $\lambda_\text{M}=2$ $\text{km}^{-2}$, $\lambda_\text{U}=500$ $\text{km}^{-2}$, and in all cases $10\%$ of the SBSs are non-IAB backhaul-connected. Thereby, we optimize the non-IAB backhaul links distribution and, as can be seen in Fig. \ref{alg_heuristic}, the GA converges rapidly with limited iterations to the maximum coverage probability achieved by exhaustive search. This is an indication of the efficiency of the proposed method with considerably lower complexity, compared to exhaustive search. Moreover, for a given number of iterations, the GA outperforms the greedy and the Tabu algorithms, in terms of coverage probability. Note that, while the greedy algorithm is easy to implement, it may not always lead to the global optimum due to the fact that it does not consider the entire search space. Finally, note that the results of Fig. \ref{alg_heuristic} are presented for a given example channel realization for which one can run the exhaustive search in limited time\footnote{We have checked the results of Fig. \ref{alg_heuristic} for a number of channel realizations and observed the same qualitative conclusions.}. However, the effectiveness of semi-optimal algorithms is more visible when studying the average system performance over multiple channel realizations, \textcolor{black}{where running exhaustive search is not feasible within limited time.}}

\section{On the Effect of Routing}\label{routing_sec}
As demonstrated, deployment planning can compensate for stationary blockages/tree foliage. On the other hand, depending on, e.g., the height of the SBSs, the (backhaul) links may be temporally blocked by, for instance, trucks passing by. In such cases, routing can be used to reduce the coverage probability degradation. For this reason, in this section, we study the effect of routing on the performance of IAB networks \textcolor{black}{(see Section \ref{3gpp_sec} for 3GPP standardization agreements on routing).} Note that, in general, routing can be utilized not only for temporal blockages but also for load balancing in the cases with varying data traffic. In this paper, we concentrate on temporal blockage, and load balancing-based routing is out of the scope of our work. Note that, here, the network deployment is first optimized based on static blockages/tree foliage. Then, by temporal blockage we refer to the blockages that are added to the network after the deployment optimization is performed.

\begin{figure}
\centerline{\includegraphics[width=3.5in]{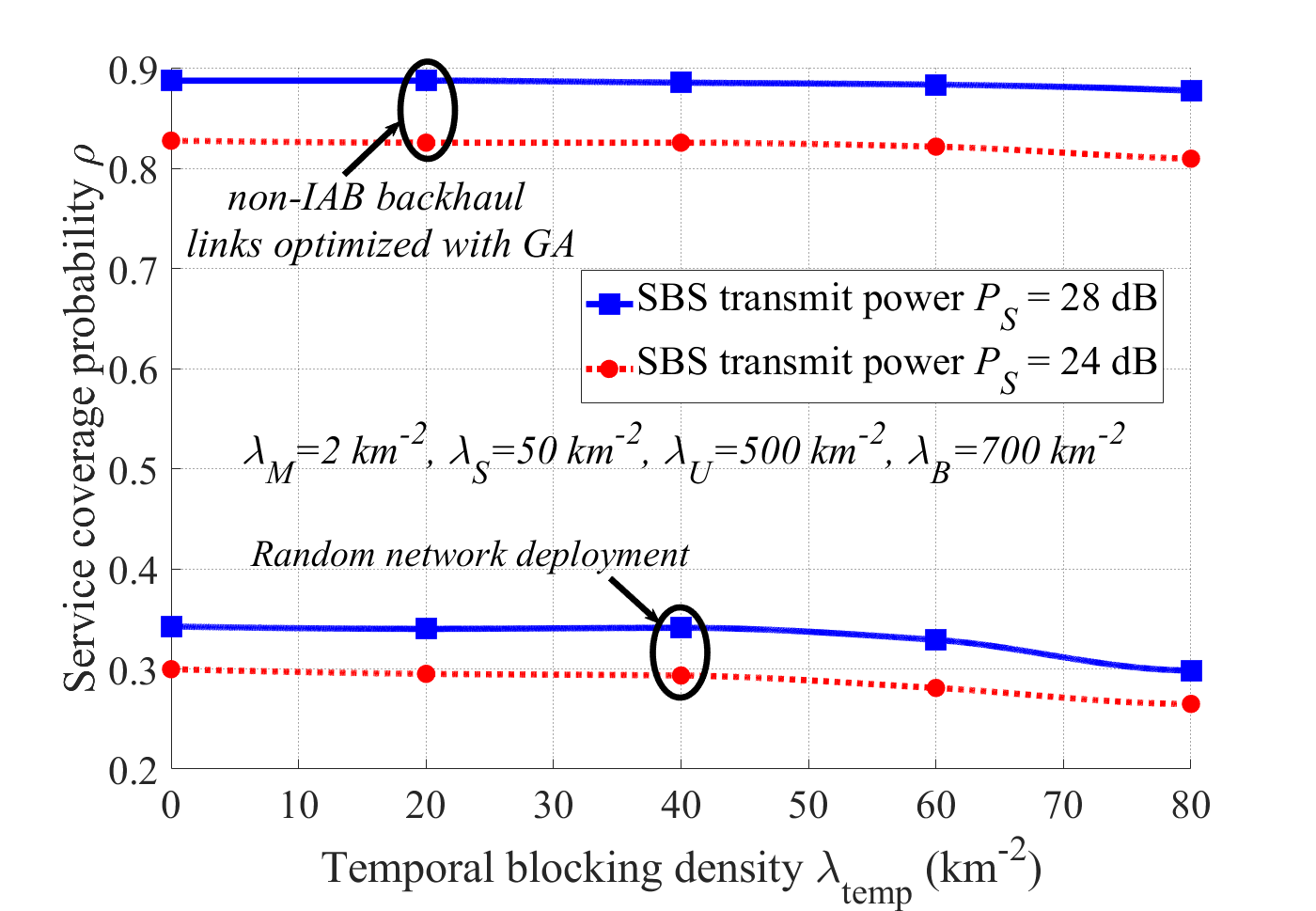}}

\caption{Service coverage probability of the IAB network as a function of the temporal blocking density $\lambda_{\text{temp}}$, with ${P_{\text{m}}, P_{\text{u}} = (40, 0)}$. The parameters are set to $\lambda_\text{M}=2$ $\text{km}^{-2}$, $\lambda_\text{S}=50$ $\text{km}^{-2}$, $\lambda_\text{U}=500$ $\text{km}^{-2}$ and $\lambda_\text{B}=700$ $\text{km}^{-2}$}  \label{temp_covg}
\end{figure}

\begin{figure}
\centerline{\includegraphics[width=3.5in]{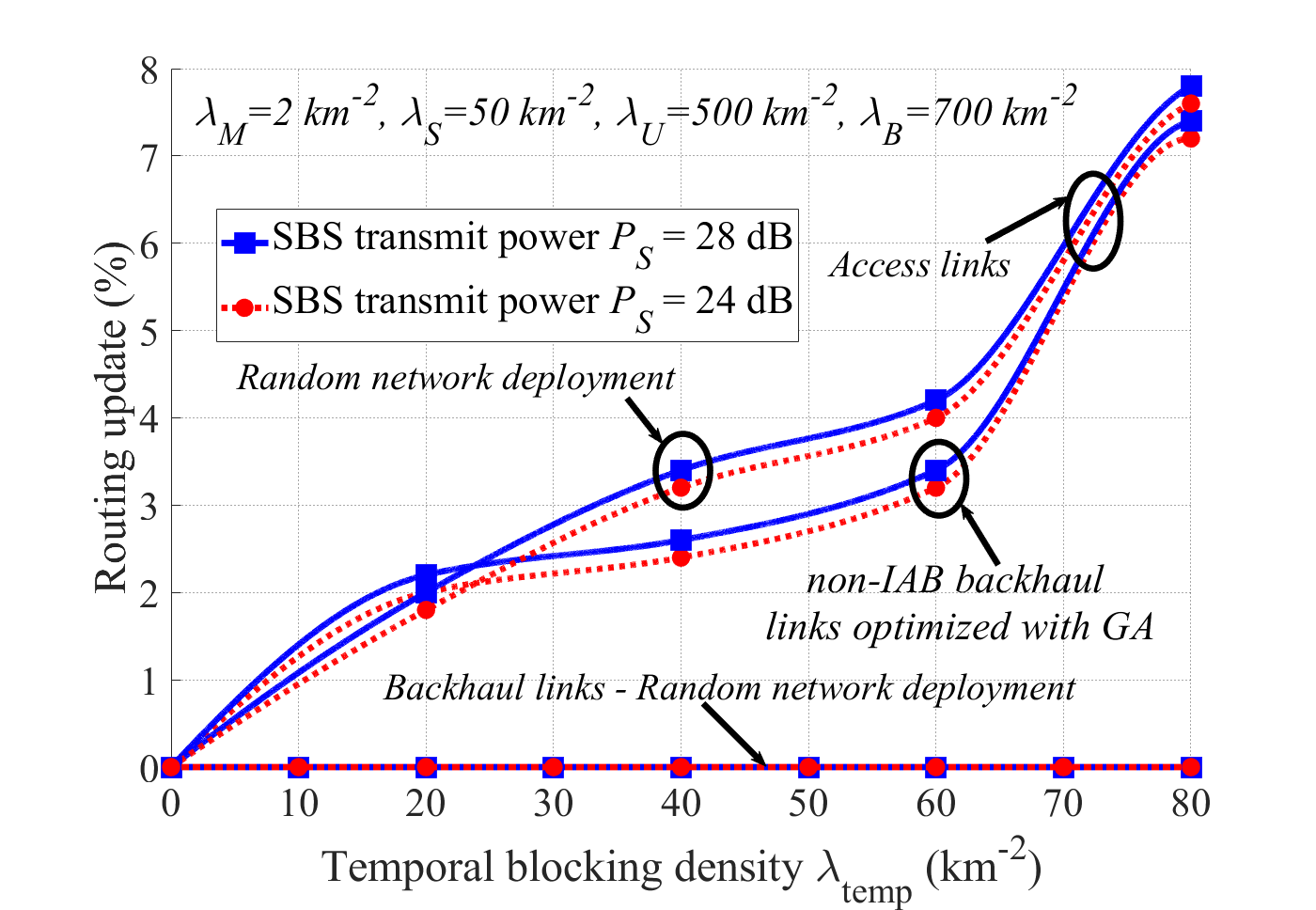}}
\caption{Percentage of routing updating of the IAB network as a function of the temporal blocking density $\lambda_{\text{temp}}$ with $P_{\text{m}}$, $P_{\text{u}} = (40, 0).$ The parameters are set to $\lambda_\text{M}=2$ $\text{km}^{-2}$, $\lambda_\text{S}=50$ $\text{km}^{-2}$, $\lambda_\text{U}=500$ $\text{km}^{-2}$ and $\lambda_\text{B}=700$ $\text{km}^{-2}$}
\label{temp_routing}
\end{figure}

\textcolor{black}{Figure  \ref{temp_covg} shows the service coverage probability considering a static blocking density $\lambda_\text{B}=700$ $\text{km}^{-2}$ for different temporal blocking densities. In addition, to understand the IAB sensitivity for temporal blockings and the effect of the routing, in Fig. \ref{temp_routing}, we plot the percentage of the links that have been updated by routing as a function of the density of the temporal blockings added to the network. The results are presented for various cases} with random deployment\textcolor{black}{, GA-optimized \textcolor{black}{non-IAB backhaul} link distribution or GA-optimized SBS locations.} Here, by routing, the received powers are recalculated, the association matrix is re-updated and thereby the data rates are calculated again, i.e., (8), (9), (13) and (14), are adapted based on the presence of temporal blockages such that the coverage probability degradation is minimized. Also, by percentage of routing update we refer to the fraction of links in the network that have changed their associated BS, both in the access and backhaul links. Here, the results are demonstrated for different transmit powers of the SBSs. According to Fig. \ref{temp_covg}-\ref{temp_routing} the following points can be concluded. 
\begin{itemize}
    \item \textcolor{black}{Unless for high densities of temporal blockings, the service coverage probability of the IAB network is not degraded much by temporal blockage (Fig. \ref{temp_covg}). Also, the introduction of GA to optimize the dedicated non-IAB backhaul connections has brought further resilience in the network to temporal blockage. Finally, the sensitivity to temporal blockage increases slightly at low SBS transmit powers (Fig. \ref{temp_covg}).}
    \item As demonstrated in Fig. \ref{temp_routing}, with temporal blockage, the routing scheme may update the access links of the UEs to the IAB nodes. However, 1) for a broad of temporal blockage densities for both \textcolor{black}{non-IAB backhaul connection-optimized} and random deployments, the access links updates are less than $10\%$. Also, 2) in all considered cases, the backhaul links do not need to be updated due to temporal blockage. This is intuitively because the IAB donor-IAB backhaul links are strong to support the required rates and the presence of temporal blockage does not affect their efficiency much unless for high temporal blockage densities. Finally, compared to random network deployment, optimizing the \textcolor{black}{non-IAB backhaul link} distribution among $10\%$ of the SBSs with GA has slightly reduced the percentage of routing update with the addition of temporal blockings. For instance, with the parameter settings of Fig. \ref{temp_routing}, $\lambda_{\text{temp}}$ = 50 $\text{km}^{-2}$, and $P_{\text{s}}=28$ dBm, in random network deployment one may need a routing update of 3.6\%, while in the GA-optimized network it is only 2.9\%.
\end{itemize}

In this way, the results indicate that, while deployment optimization can well robustify the network to static blockages, with a well-planned network the system performance is almost insensitive to low/moderate temporal blockages, and routing may not be required unless for high temporal blockage densities/severe coverage probability requirements. On the other hand, depending on the data traffic variation and the number of hops in the IAB network, the routing may be of interest in load balancing.

\section{Conclusion}\label{conclusion}
We studied the problem of deployment optimization and routing in IAB networks to guarantee high coverage probability in the presence of tree foliage/blockage. Moreover, we reviewed the recent 3GPP agreements on IAB-based routing, as well as the key challenges to enable meshed IAB. 

As we showed, machine-learning techniques can be effectively utilized for deployment optimization, with no need for mathematical analysis and with the capability to be adapted for different channel models/constraints/metrics of interest. \textcolor{black}{Particularly, the proposed algorithm reduces the complexity compared to exhaustive search significantly because with typical network area our proposed scheme requires orders of magnitude less solution checkings compared to exhaustive search. Also, while deployment planning boosts the coverage probability of IAB networks, compared to random deployment, significantly, for a broad range of coverage constraints/blockage densities, the impact of routing to increase redundancy may be negligible.} Indeed, routing may be of interest in the cases with severe availability constraints/high blockage densities as well as for load balancing. Finally, in practice, deployment planning may be affected by, e.g., the availability of \textcolor{black}{non-IAB backhaul connection} in specific areas, and the designer may consider, e.g., seasonal tree foliage variations, rental costs and/or foreseen infrastructure changes.

\textcolor{black}{
\section{Acknowledgment}
This work was supported in part by VINNOVA (Swedish Government Agency for Innovation Systems) within the VINN Excellence Center ChaseOn and in part by the European Commission through the H2020 project Hexa-X (Grant Agreement no. 101015956).}

\bibliographystyle{IEEEtran}
\bibliography{bibliography}

\begin{thebibliography}{10}
\providecommand{\url}[1]{#1}
\csname url@samestyle\endcsname
\providecommand{\newblock}{\relax}
\providecommand{\bibinfo}[2]{#2}
\providecommand{\BIBentrySTDinterwordspacing}{\spaceskip=0pt\relax}
\providecommand{\BIBentryALTinterwordstretchfactor}{4}
\providecommand{\BIBentryALTinterwordspacing}{\spaceskip=\fontdimen2\font plus
\BIBentryALTinterwordstretchfactor\fontdimen3\font minus
  \fontdimen4\font\relax}
\providecommand{\BIBforeignlanguage}[2]{{%
\expandafter\ifx\csname l@#1\endcsname\relax
\typeout{** WARNING: IEEEtran.bst: No hyphenation pattern has been}%
\typeout{** loaded for the language `#1'. Using the pattern for}%
\typeout{** the default language instead.}%
\else
\language=\csname l@#1\endcsname
\fi
#2}}
\providecommand{\BIBdecl}{\relax}
\BIBdecl

\bibitem{eref1}
\BIBentryALTinterwordspacing
P.~{Cerwall}, ``Mobile data traffic growth outlook,'' Jun. 2018. [Online].
  Available: \url{https://www.ericsson.com/en/mobility-report/reports}
\BIBentrySTDinterwordspacing

\bibitem{eref2}
\BIBentryALTinterwordspacing
Ericsson, ``Mobile data traffic growth outlook.'' [Online]. Available:
  \url{https://www.ericsson.com/en/mobility-report/reports/november-2018/mobile-data-traffic-growth-outlook}
\BIBentrySTDinterwordspacing

\bibitem{eref3}
M.~{Agiwal}, A.~{Roy}, and N.~{Saxena}, ``{Next Generation 5G Wireless
  Networks: A Comprehensive Survey},'' \emph{Commun. Surv. Tutor.}, vol.~18,
  no.~3, pp. 1617--1655, Feb. 2016.

\bibitem{eref4}
S.~Dang, O.~Amin, B.~Shihada, and M.-S. Alouini, ``{What should 6G be?}''
  \emph{Nature Electronics}, vol.~3, no.~1, pp. 20--29, Jan. 2020.

\bibitem{eref5}
X.~Gao, O.~Edfors, F.~Tufvesson, and E.~G. Larsson, ``Massive mimo in real
  propagation environments: Do all antennas contribute equally?'' \emph{IEEE
  Trans. Wirel. Commun.}, vol.~63, no.~11, pp. 3917--3928, July 2015.

\bibitem{eref7}
C.~{Czegledi}, M.~{Hörberg}, M.~{Sjödin}, P.~{Ligander}, J.~{Hansryd},
  J.~{Sandberg}, J.~{Gustavsson}, D.~{Sjoöberg}, D.~{Polydorou}, and
  D.~{Siomos}, ``{Demonstrating 139 Gbps and 55.6 bps/Hz Spectrum Efficiency
  Using 8x8 MIMO over a 1.5 km Link at 73.5 GHz},'' accepted for presentation
  in IMS2020, CA, USA, Aug. 2020.

\bibitem{eref8}
C.~{Dehos}, J.~L. {González}, A.~D. {Domenico}, D.~{Kténas}, and
  L.~{Dussopt}, ``{Millimeter-wave access and backhauling: the solution to the
  exponential data traffic increase in 5G mobile communications systems?}''
  \emph{IEEE Commun. Mag.}, vol.~52, no.~9, pp. 88--95, Sept. 2014.

\bibitem{eref9}
Y.~{Li}, E.~{Pateromichelakis}, N.~{Vucic}, J.~{Luo}, W.~{Xu}, and G.~{Caire},
  ``{Radio Resource Management Considerations for 5G Millimeter Wave Backhaul
  and Access Networks},'' \emph{IEEE Commun. Mag.}, vol.~55, no.~6, pp. 86--92,
  Jun. 2017.

\bibitem{terabit_surv}
N.~Rajatheva, I.~Atzeni, S.~Bicais, E.~Bjornson, A.~Bourdoux, S.~Buzzi,
  C.~D'Andrea, J.-B. Dore, S.~Erkucuk, M.~Fuentes \emph{et~al.}, ``{Scoring the
  terabit/s goal: Broadband connectivity in 6G},'' \emph{arXiv preprint
  arXiv:2008.07220}, 2020.

\bibitem{eref27}
K.~N. R. S.~V. {Prasad}, H.~K. {Rath}, and A.~{Simha}, ``Wireless mobile
  network planning and optimization,'' in \emph{Proc. IEEE COMSNETS'2014},
  Bangalore, India, 2014, pp. 1--4.

\bibitem{eref28}
A.~{Al-Dulaimi}, S.~{Al-Rubaye}, J.~{Cosmas}, and A.~{Anpalagan}, ``Planning of
  ultra-dense wireless networks,'' \emph{IEEE Network}, vol.~31, no.~2, pp.
  90--96, March/April 2017.

\bibitem{eref10}
M.~N. {Islam}, S.~{Subramanian}, and A.~{Sampath}, ``{Integrated Access
  Backhaul in Millimeter Wave Networks},'' in \emph{Proc. IEEE WCNC'2017}, CA,
  USA, Mar. 2017, pp. 1--6.

\bibitem{eref11}
M.~N. {Islam}, N.~{Abedini}, G.~{Hampel}, S.~{Subramanian}, and J.~{Li},
  ``{Investigation of performance in integrated access and backhaul
  networks},'' in \emph{Proc. IEEE INFOCOM WKSHPS'2018}, HI, USA, Apr. 2018,
  pp. 597--602.

\bibitem{eref12}
\BIBentryALTinterwordspacing
Y.~Li, J.~Luo, R.~A. Stirling-Gallacher, and G.~Caire, ``{Integrated access and
  backhaul optimization for millimeter wave heterogeneous networks},''
  \emph{arXiv}, 2019. [Online]. Available:
  \url{https://arxiv.org/abs/1901.04959.}
\BIBentrySTDinterwordspacing

\bibitem{eref13}
Y.~{Liu}, A.~{Tang}, and X.~{Wang}, ``{Joint Incentive and Resource Allocation
  Design for User Provided Network Under 5G Integrated Access and Backhaul
  Networks},'' \emph{IEEE Trans. Netw. Sci. Eng.}, vol.~7, no.~2, pp. 673--685,
  April-June 2020.

\bibitem{eref14}
N.~{Tafintsev}, D.~{Moltchanov}, M.~{Simsek}, S.~{Yeh}, S.~{Andreev},
  Y.~{Koucheryavy}, and M.~{Valkama}, ``Reinforcement learning for improved
  uav-based integrated access and backhaul operation,'' in \emph{Proc. IEEE ICC
  Workshops'2017}, 2020, pp. 1--7.

\bibitem{eref15}
M.~N. {Kulkarni}, J.~G. {Andrews}, and A.~{Ghosh}, ``{Performance of Dynamic
  and Static TDD in Self-Backhauled Millimeter Wave Cellular Networks},''
  \emph{IEEE Trans. Wireless Commun.}, vol.~16, no.~10, pp. 6460--6478, Oct.
  2017.

\bibitem{eref16}
B.~{Makki}, M.~{Hashemi}, L.~{Bao}, and M.~{Coldrey}, ``{On the Performance of
  FDD and TDD Systems in Different Data Traffics: Finite Block-Length
  Analysis},'' in \emph{Proc. VTC-Fall'2018}, IL, USA, Aug. 2018, pp. 1--5.

\bibitem{eref17}
C.~{Saha}, M.~{Afshang}, and H.~S. {Dhillon}, ``{Integrated mmWave Access and
  Backhaul in 5G: Bandwidth Partitioning and Downlink Analysis},'' in
  \emph{Proc. ICC'2018}, MO, USA, May 2018, pp. 1--6.

\bibitem{eref18}
S.~{Singh}, M.~N. {Kulkarni}, A.~{Ghosh}, and J.~G. {Andrews}, ``{Tractable
  Model for Rate in Self-Backhauled Millimeter Wave Cellular Networks},''
  \emph{IEEE J. Sel. Areas Commun.}, vol.~33, no.~10, pp. 2196--2211, Oct.
  2015.

\bibitem{eref19}
\BIBentryALTinterwordspacing
C.~{Saha} and H.~S. {Dhillon}, ``{Millimeter Wave Integrated Access and
  Backhaul in 5G: Performance Analysis and Design Insights},'' \emph{arXiv},
  2019. [Online]. Available: \url{https://arxiv.org/pdf/1902.06300.pdf}
\BIBentrySTDinterwordspacing

\bibitem{eref20}
M.~{Polese}, M.~{Giordani}, A.~{Roy}, S.~{Goyal}, D.~{Castor}, and M.~{Zorzi},
  ``{End-to-End Simulation of Integrated Access and Backhaul at mmWaves},'' in
  \emph{Proc. IEEE CAMAD'2018}, Barcelona, Spain, Sept. 2018, pp. 1--7.

\bibitem{eref21}
M.~{Polese}, M.~{Giordani}, T.~{Zugno}, A.~{Roy}, S.~{Goyal}, D.~{Castor}, and
  M.~{Zorzi}, ``{Integrated Access and Backhaul in 5G mmWave Networks:
  Potential and Challenges},'' \emph{IEEE Commun. Mag.}, vol.~58, no.~3, pp.
  62--68, Mar. 2020.

\bibitem{eref22}
M.~{Hashemi}, M.~{Coldrey}, M.~{Johansson}, and S.~{Petersson}, ``{Integrated
  Access and Backhaul in Fixed Wireless Access Systems},'' in \emph{Proc.
  VTC-Fall'2017}, Toronto, Canada, Sept. 2017, pp. 1--5.

\bibitem{eref23}
N.~{Tafintsev}, D.~{Moltchanov}, M.~{Gerasimenko}, M.~{Gapeyenko}, J.~{Zhu},
  S.~{Yeh}, N.~{Himayat}, S.~{Andreev}, Y.~{Koucheryavy}, and M.~{Valkama},
  ``{Aerial Access and Backhaul in mmWave B5G Systems: Performance Dynamics and
  Optimization},'' \emph{IEEE Commun. Mag.}, vol.~58, no.~2, pp. 93--99, Feb.
  2020.

\bibitem{eref24}
O.~{Teyeb}, A.~{Muhammad}, G.~{Mildh}, E.~{Dahlman}, F.~{Barac}, and
  B.~{Makki}, ``{Integrated Access Backhauled Networks},'' in \emph{Proc.
  VTC2019-Fall'2019}, Honolulu, HI, USA, Sept. 2019, pp. 1--5.

\bibitem{erefour}
C.~{Madapatha}, B.~{Makki}, C.~{Fang}, O.~{Teyeb}, E.~{Dahlman}, M.~{Alouini},
  and T.~{Svensson}, ``On integrated access and backhaul networks: Current
  status and potentials,'' \emph{IEEE Open Journal of the Communications
  Society}, pp. 1--1, Sept. 2020.

\bibitem{eref25}
F.~{Gómez-Cuba} and M.~{Zorzi}, ``Twice simulated annealing resource
  allocation for mmwave multi-hop networks with interference.'' in \emph{Proc.
  IEEE ICC'2020}, Dublin, Ireland, 2020, pp. 1--7.

\bibitem{eref26}
A.~{Fouda}, A.~S. {Ibrahim}, I.~{Guvenc}, and M.~{Ghosh}, ``{UAV-Based In-Band
  Integrated Access and Backhaul for 5G Communications},'' in \emph{Proc. IEEE
  VTC-Fall'2018}, IL, USA, Aug. 2018, pp. 1--5.

\bibitem{eref30}
A.~{HasanzadeZonuzy}, I.~{Hou}, and S.~{Shakkottai}, ``{Broadcasting Real-Time
  Flows in Integrated Backhaul and Access 5G Networks},'' in \emph{Proc. IEEE
  WiOPT'2019}, Avignon, France, 2019, pp. 1--8.

\bibitem{eref31}
M.~{Polese}, M.~{Giordani}, A.~{Roy}, D.~{Castor}, and M.~{Zorzi},
  ``{Distributed Path Selection Strategies for Integrated Access and Backhaul
  at mmWaves},'' in \emph{Proc. IEEE GLOBECOM'2018}, Abu Dhabi, United Arab
  Emirates, Dec. 2018, pp. 1--7.

\bibitem{eref32}
B.~{Zhai}, M.~{Yu}, A.~{Tang}, and X.~{Wang}, ``Mesh architecture for efficient
  integrated access and backhaul networking,'' in \emph{Proc. IEEE WCNC'2020},
  Seoul, Korea (South), May 2020, pp. 1--6.

\bibitem{eref33}
J.~Y. {Lai}, W.~{Wu}, and Y.~T. {Su}, ``Resource allocation and node placement
  in multi-hop heterogeneous integrated-access-and-backhaul networks,''
  \emph{IEEE Access}, vol.~8, pp. 122\,937--122\,958, July 2020.

\bibitem{gen4}
N.~Chen, T.~Qiu, X.~Zhou, K.~Li, and M.~Atiquzzaman, ``An intelligent robust
  networking mechanism for the internet of things,'' \emph{IEEE Commun. Mag.},
  vol.~57, no.~11, pp. 91--95, Nov. 2019.

\bibitem{gen5}
X.~Meng, H.~Inaltekin, and B.~Krongold, ``Deep reinforcement learning-based
  topology optimization for self-organized wireless sensor networks,'' in
  \emph{Proc. IEEE GLOBECOM'2019}.\hskip 1em plus 0.5em minus 0.4em\relax HI,
  USA: IEEE, 2019, pp. 1--6.

\bibitem{gen6}
X.~Fu, F.~R. Yu, J.~Wang, Q.~Qi, and J.~Liao, ``Dynamic service function chain
  embedding for nfv-enabled iot: A deep reinforcement learning approach,''
  \emph{IEEE Trans. Wirel. Commun.}, vol.~19, no.~1, pp. 507--519, Oct. 2019.

\bibitem{gen7}
P.~Sun, Y.~Hu, J.~Lan, L.~Tian, and M.~Chen, ``Tide: Time-relevant deep
  reinforcement learning for routing optimization,'' \emph{Future Generation
  Computer Systems}, vol.~99, pp. 401--409, Oct. 2019.

\bibitem{gen8}
M.~Wang, Y.~Cui, S.~Xiao, X.~Wang, D.~Yang, K.~Chen, and J.~Zhu, ``Neural
  network meets dcn: Traffic-driven topology adaptation with deep learning,''
  \emph{Proceedings of the ACM on Measurement and Analysis of Computing
  Systems}, vol.~2, no.~2, pp. 1--25, Jun. 2018.

\bibitem{gen9}
S.~Zhang, B.~Yin, and Y.~Cheng, ``Topology aware deep learning for wireless
  network optimization,'' \emph{arXiv preprint arXiv:1912.08336}, 2019.

\bibitem{gen10}
P.~V. {Klaine}, M.~A. {Imran}, O.~{Onireti}, and R.~D. {Souza}, ``A survey of
  machine learning techniques applied to self-organizing cellular networks,''
  \emph{Commun. Surv. Tutor.}, vol.~19, no.~4, pp. 2392--2431, July 2017.

\bibitem{gen11}
Q.~{Mao}, F.~{Hu}, and Q.~{Hao}, ``Deep learning for intelligent wireless
  networks: A comprehensive survey,'' \emph{Commun. Surv. Tutor.}, vol.~20,
  no.~4, pp. 2595--2621, Jun. 2018.

\bibitem{gen12}
S.~Troia, R.~Alvizu, and G.~Maier, ``Reinforcement learning for service
  function chain reconfiguration in nfv-sdn metro-core optical networks,''
  \emph{IEEE Access}, vol.~7, pp. 167\,944--167\,957, Nov. 2019.

\bibitem{gen13}
S.~{Wang} and T.~{Lv}, ``Deep reinforcement learning for demand-aware joint vnf
  placement-and-routing,'' in \emph{Proc. IEEE GC Wkshps'2019}, HI, USA, 2019,
  pp. 1--6.

\bibitem{gen1}
X.~Du, H.~Van~Nguyen, C.~Jiang, Y.~Li, F.~R. Yu, and Z.~Han, ``Virtual relay
  selection in lte-v: A deep reinforcement learning approach to heterogeneous
  data,'' \emph{IEEE Access}, May 2020.

\bibitem{gen2}
S.~Yan, X.~Zhang, H.~Xiang, and W.~Wu, ``Joint access mode selection and
  spectrum allocation for fog computing based vehicular networks,'' \emph{IEEE
  Access}, vol.~7, pp. 17\,725--17\,735, Jan. 2019.

\bibitem{gen3}
Y.~Chang, X.~Yuan, B.~Li, D.~Niyato, and N.~Al-Dhahir, ``Machine-learning-based
  parallel genetic algorithms for multi-objective optimization in
  ultra-reliable low-latency wsns,'' \emph{IEEE Access}, vol.~7, pp.
  4913--4926, Dec. 2018.

\bibitem{gen14}
Y.~{Xu}, W.~{Xu}, Z.~{Wang}, J.~{Lin}, and S.~{Cui}, ``Load balancing for
  ultradense networks: A deep reinforcement learning-based approach,''
  \emph{IEEE Internet Things J.}, vol.~6, no.~6, pp. 9399--9412, Aug. 2019.

\bibitem{gen15}
X.~{Chen}, R.~{Proietti}, H.~{Lu}, A.~{Castro}, and S.~J.~B. {Yoo},
  ``Knowledge-based autonomous service provisioning in multi-domain elastic
  optical networks,'' \emph{IEEE Commun. Mag.}, vol.~56, no.~8, pp. 152--158,
  Aug. 2018.

\bibitem{eref29}
S.~M. {Azimi-Abarghouyi}, B.~{Makki}, M.~{Haenggi}, M.~{Nasiri-Kenari}, and
  T.~{Svensson}, ``{Coverage analysis of finite cellular networks: A stochastic
  geometry approach},'' in \emph{Proc. IEEE IWCIT'2018}, Tehran, Iran, Apr.
  2018, pp. 1--5.

\bibitem{3gpp_rp201756}
{3GPP RP-201756}, ``{Revised WID: Integrated Access and Backhaul for NR},''
  {3rd Generation Partnership Project (3GPP)}, Tech. Rep. Meeting RAN\#89e,
  Electronic Meeting, Jun. 2020.

\bibitem{3gpp_rp201293}
{3GPP RP-201293}, ``{New WID on Enhancements to Integrated Access and Backhaul
  },'' {3rd Generation Partnership Project (3GPP)}, Tech. Rep. Meeting
  RAN\#88-e, Electronic Meeting, Sep. 2020.

\bibitem{g1}
T.~K. Vu, M.~Bennis, S.~Samarakoon, M.~Debbah, and M.~Latva-aho, ``{Joint Load
  Balancing and Interference Mitigation in 5G Heterogeneous Networks},''
  \emph{IEEE Transactions on Wireless Communications}, vol.~16, no.~9, pp.
  6032--6046, 2017.

\bibitem{g2}
T.~K. Vu, ``{Integrated access-backhaul for 5G wireless network},''
  \emph{University of Oulu}, 2019.

\bibitem{gensys2}
\BIBentryALTinterwordspacing
M.~N. Kulkarni, A.~Ghosh, and J.~G. Andrews, ``{Max-min rates in
  self-backhauled millimeter wave cellular networks},'' \emph{arXiv}, 2018.
  [Online]. Available: \url{https://arxiv.org/abs/1805.01040}
\BIBentrySTDinterwordspacing

\bibitem{gensys3}
F.~{Gomez-Cuba} and M.~{Zorzi}, ``Optimal link scheduling in millimeter wave
  multi-hop networks with space division multiple access,'' in \emph{Proc. IEEE
  ITA'2016}, La Jolla, CA, Feb. 2016, pp. 1--9.

\bibitem{gensys4}
J.~Y. {Lai}, W.~{Wu}, and Y.~T. {Su}, ``Resource allocation and node placement
  in multi-hop heterogeneous integrated-access-and-backhaul networks,'' vol.~8,
  Jul. 2020, pp. 122\,937--122\,958.

\bibitem{g3}
\BIBentryALTinterwordspacing
H.~Ronkainen, J.~Edstam, C.~{\"O}stberg, and A.~Ericsson, ``{Integrated access
  and backhaul – a new type of wireless backhaul in 5G},'' Jun. 2020.
  [Online]. Available:
  \url{https://www.ericsson.com/en/reports-and-papers/ericsson-technology-review/articles/introducing-integrated-access-and-backhaul}
\BIBentrySTDinterwordspacing

\bibitem{ref5}
\BIBentryALTinterwordspacing
B.~B{\l}aszczyszyn, ``{Lecture Notes on Random Geometric Models---Random
  Graphs, Point Processes and Stochastic Geometry},'' 2017. [Online].
  Available: \url{https://hal.inria.fr/cel-01654766/document}
\BIBentrySTDinterwordspacing

\bibitem{ref1}
S.~M. {Azimi-Abarghouyi}, B.~{Makki}, M.~{Nasiri-Kenari}, and T.~{Svensson},
  ``{Stochastic Geometry Modeling and Analysis of Finite Millimeter Wave
  Wireless Networks},'' \emph{IEEE Trans. Veh. Technol.}, vol.~68, no.~2, pp.
  1378--1393, Feb. 2019.

\bibitem{ref3}
T.~S. {Rappaport}, Y.~{Xing}, G.~R. {MacCartney}, A.~F. {Molisch},
  E.~{Mellios}, and J.~{Zhang}, ``{Overview of Millimeter Wave Communications
  for Fifth-Generation (5G) Wireless Networks—With a Focus on Propagation
  Models},'' \emph{IEEE Trans. Antennas Propag.}, vol.~65, no.~12, pp.
  6213--6230, Dec. 2017.

\bibitem{gensys6}
P.~J. Diggle, \emph{{Some statistical aspects of spatial distribution models
  for plants and trees}}, 1982, no. 162.

\bibitem{gensys7}
M.~A. Abu-Rgheff, \emph{{5G Physical Layer Technologies}}.\hskip 1em plus 0.5em
  minus 0.4em\relax John Wiley \& Sons, Nov. 2019, pp. 303-313.

\bibitem{genb1}
B.~{Makki}, A.~{Ide}, T.~{Svensson}, T.~{Eriksson}, and M.~{Alouini}, ``A
  genetic algorithm-based antenna selection approach for large-but-finite mimo
  networks,'' \emph{IEEE Trans. Veh. Technol.}, vol.~66, no.~7, pp. 6591--6595,
  Dec. 2017.

\bibitem{nphard1}
\BIBentryALTinterwordspacing
{Tutorialspoint}, ``{NP} hard and {NP}-complete classes,'' Accessed Aug. 20,
  2021. [Online]. Available:
  \url{https://www.tutorialspoint.com/design_and_analysis_of_algorithms/design_and_analysis_of_algorithms_np_hard_complete_classes.htm}
\BIBentrySTDinterwordspacing

\bibitem{nphard2}
G.~Panchal and D.~Panchal, ``Solving {NP} hard problems using genetic
  algorithm,'' \emph{Transportation}, vol. 106, pp. 6--2, 2015.

\bibitem{steinmetzer2017compressive}
D.~Steinmetzer, D.~Wegemer, M.~Schulz, J.~Widmer, and M.~Hollick, ``Compressive
  millimeter-wave sector selection in off-the-shelf ieee 802.11 ad devices,''
  in \emph{Proc. IEEE CoNEXT'2017}, NY, USA, Nov. 2017, pp. 414--425.

\bibitem{erefaddnewc}
\BIBentryALTinterwordspacing
H.~{Ronkainen}, J.~{Edstam}, A.~{ Ericsson}, and C.~{Östberg}, ``Ericsson
  technology review.'' [Online]. Available:
  \url{https://www.ericsson.com/en/reports-and-papers/ericsson-technology-review/articles/introducing-integrated-access-and-backhaul}
\BIBentrySTDinterwordspacing

\bibitem{z_1}
H.~Guo, B.~Makki, and T.~Svensson, ``Genetic algorithm-based beam refinement
  for initial access in millimeter wave mobile networks,'' \emph{Wirel. Commun.
  Mob. Comput.}, vol. 2018, Jun. 2018.

\bibitem{z_3}
Z.~Gu, J.~Zhang, and Y.~Ji, ``{Topology optimization for FSO-based
  fronthaul/backhaul in 5G+ wireless networks},'' in \emph{Proc. IEEE ICC
  Workshops'2018}, MO, USA, May 2018, pp. 1--6.

\bibitem{z_2}
X.~Gao, L.~Dai, C.~Yuen, and Z.~Wang, ``Turbo-like beamforming based on tabu
  search algorithm for millimeter-wave massive mimo systems,'' \emph{IEEE
  Trans. Veh. Technol.}, vol.~65, no.~7, pp. 5731--5737, 2015.

\end{thebibliography}

\begin{IEEEbiography}[{\includegraphics[width=1in,clip,keepaspectratio]{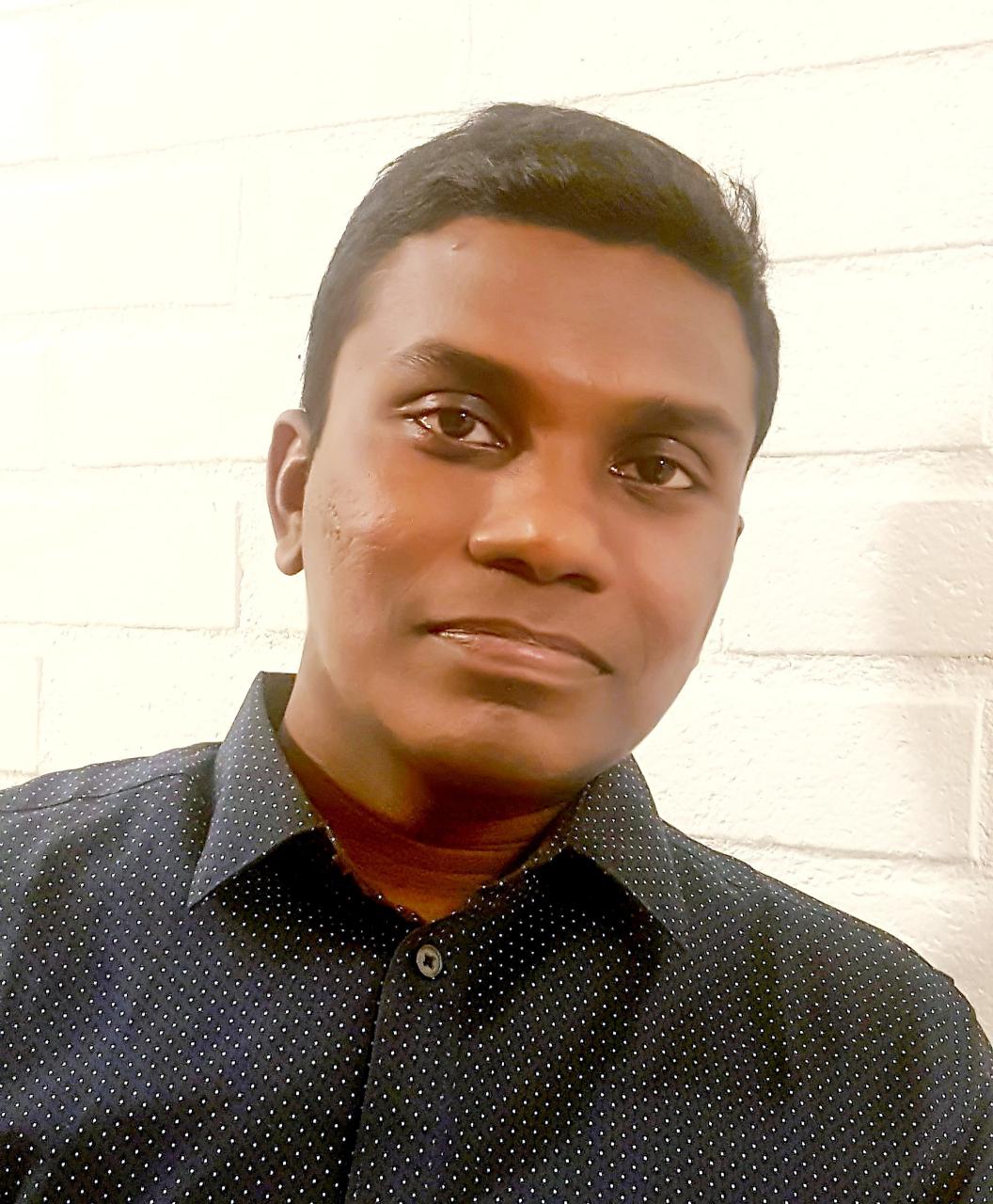}}]{Charitha Madapatha} \textcolor{black}{is currently pursuing the Ph.D. in Communication Systems and Information Theory from Chalmers University of Technology, Gothenburg, Sweden. He received his MSc. degree in Communication Engineering from Chalmers in 2020 and the BSc. degree in Telecommuncations Engineering from Asian Institute of Technology, Pathumthani, Thailand, in 2016. }

Charitha is the recipient of Swedish Institute Scholarship for Global Professionals, Sweden, 2018 and the AIT Fellowship grant, Thailand, 2014-2016. He was also involved in LTE network planning and optimization in Thailand and Cambodia. His current research interests include integrated access and backhaul, multi antenna systems, mmWave communication, analysis of physical layer algorithms, resource allocation. He has co-authored several international scientific publications in the field of wireless networks.

\end{IEEEbiography}

\begin{IEEEbiography}[{\includegraphics[width=1in,clip,keepaspectratio]{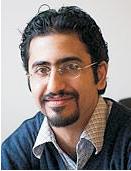}}]{Behrooz Makki} [M'19, SM'19] received his PhD degree in Communication Engineering from Chalmers University of Technology, Gothenburg, Sweden. In 2013-2017, he was a Postdoc researcher at Chalmers University. Currently, he works as Senior Researcher in Ericsson Research, Gothenburg, Sweden.

Behrooz is the recipient of the VR Research Link grant, Sweden, 2014, the Ericsson’s Research grant, Sweden, 2013, 2014 and 2015, the ICT SEED grant, Sweden, 2017, as well as the Wallenbergs research grant, Sweden, 2018. Also, Behrooz is the recipient of the IEEE best reviewer award, IEEE Transactions on Wireless Communications, 2018, and the IEEE best Editor award, IEEE Wireless Communications Letters, 2020. Currently, he works as an Editor in IEEE Wireless Communications Letters, IEEE Communications Letters, the journal of Communications and Information Networks, as well as the Associate Editor in Frontiers in Communications and Networks.

He was a member of European Commission projects “mm-Wave based Mobile Radio Access Network for 5G Integrated Communications” and “ARTIST4G” as well as various national and international research collaborations. His current research interests include integrated access and backhaul, hybrid automatic repeat request, Green communications, millimeter wave communications, free-space optical communication, NOMA, finite block-length analysis and backhauling. He has co-authored 67 journal papers, 47 conference papers and 70 patent applications.

\end{IEEEbiography}

\begin{IEEEbiography}[{\includegraphics[width=1in,clip,keepaspectratio]{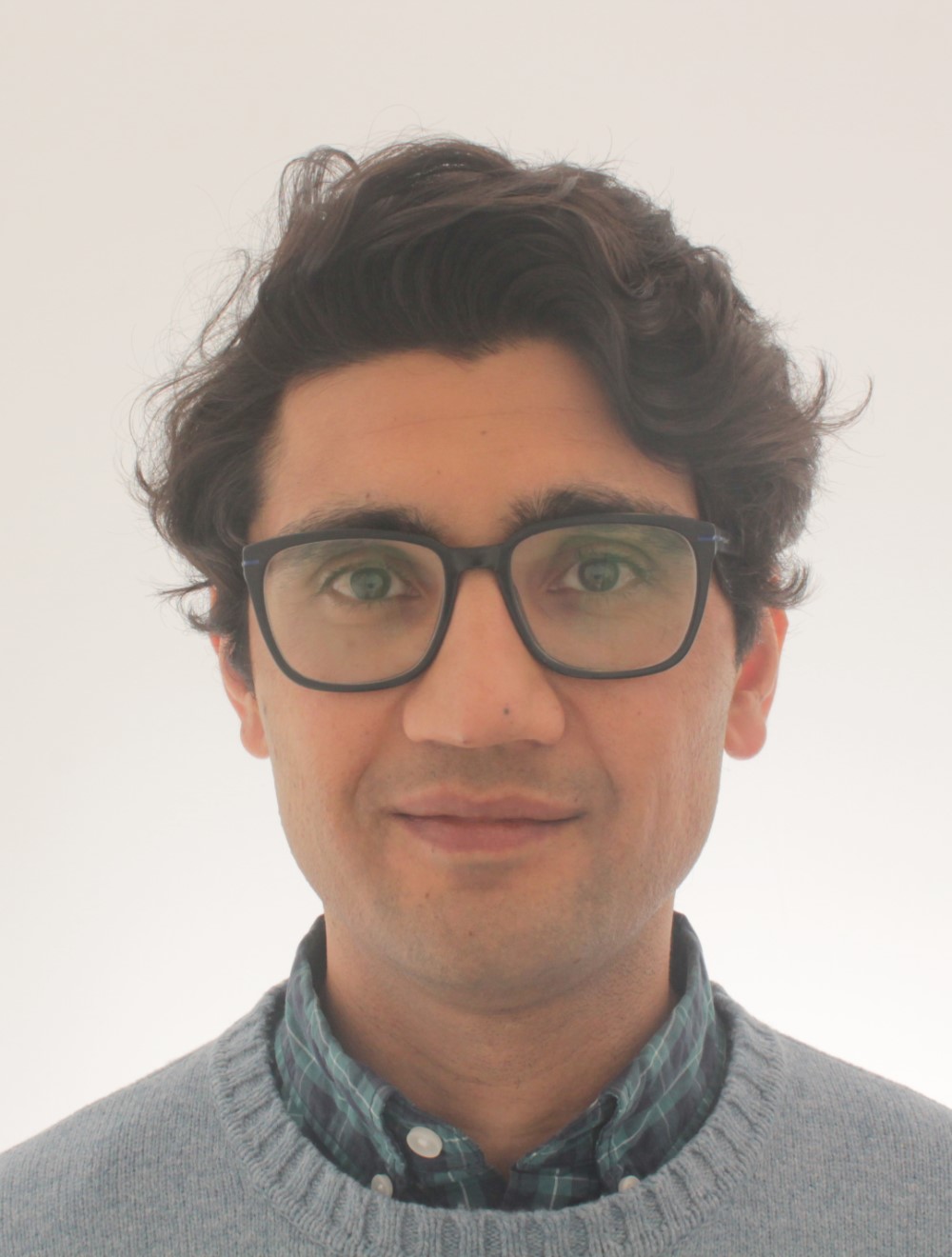}}]{Ajmal Muhammad} received his PhD degree in Information Coding from Linköping University, Linköping, Sweden in 2015. During 2015-2017, he worked as a PostDoc researcher at ONLab, KTH-Royal Institute of Technology, Stockholm, Sweden. Currently, he is working as Experienced Researcher at Ericsson Research, Kista. Ajmal has co-authored 41 research publications including journal and conference papers, and 30 patent applications.
\end{IEEEbiography}

\begin{IEEEbiography}[{\includegraphics[width=1in,clip,keepaspectratio]{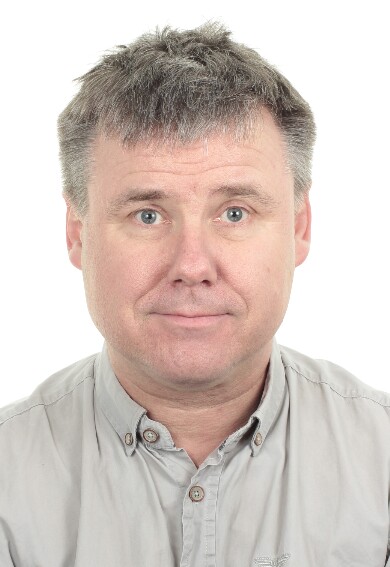}}]{Dr. Erik Dahlman} is currently Senior Expert in Radio Access Technologies within Ericsson Research. He has been deeply involved in the development of all 3GPP wireless access technologies, from the early 3G technologies (WCDMA(HSPA), via 4G LTE, and most recently the 5G NR technology. He current work is primarily focusing on the evolution of 5G as well as technologies applicable to future beyond 5G wireless access.

Erik Dahlman is the co-author of the books 3G Evolution – HSPA and LTE for Mobile Broadband, 4G – LTE and LTE-Advanced for mobile broadband, 4G – LTE-Advanced Pro and The Road to 5G and, most recently, 5G NR – The Next Generation Wireless Access Technology.

In 2009, Erik Dahlman received the Major Technical Award, an award handed out by the Swedish Government, for his contributions to the technical and commercial success of the 3G HSPA radio-access technology. In 2010, he was part of the Ericsson team receiving the LTE Award for “Best Contribution to LTE Standards”, handed out at the LTE World Summit. In 2014 he was nominated for the European Inventor Award, the most prestigious inventor award in Europe, for contributions to the development of 4G LTE.
\end{IEEEbiography}

\begin{IEEEbiography}[{\includegraphics[width=1in,clip,keepaspectratio]{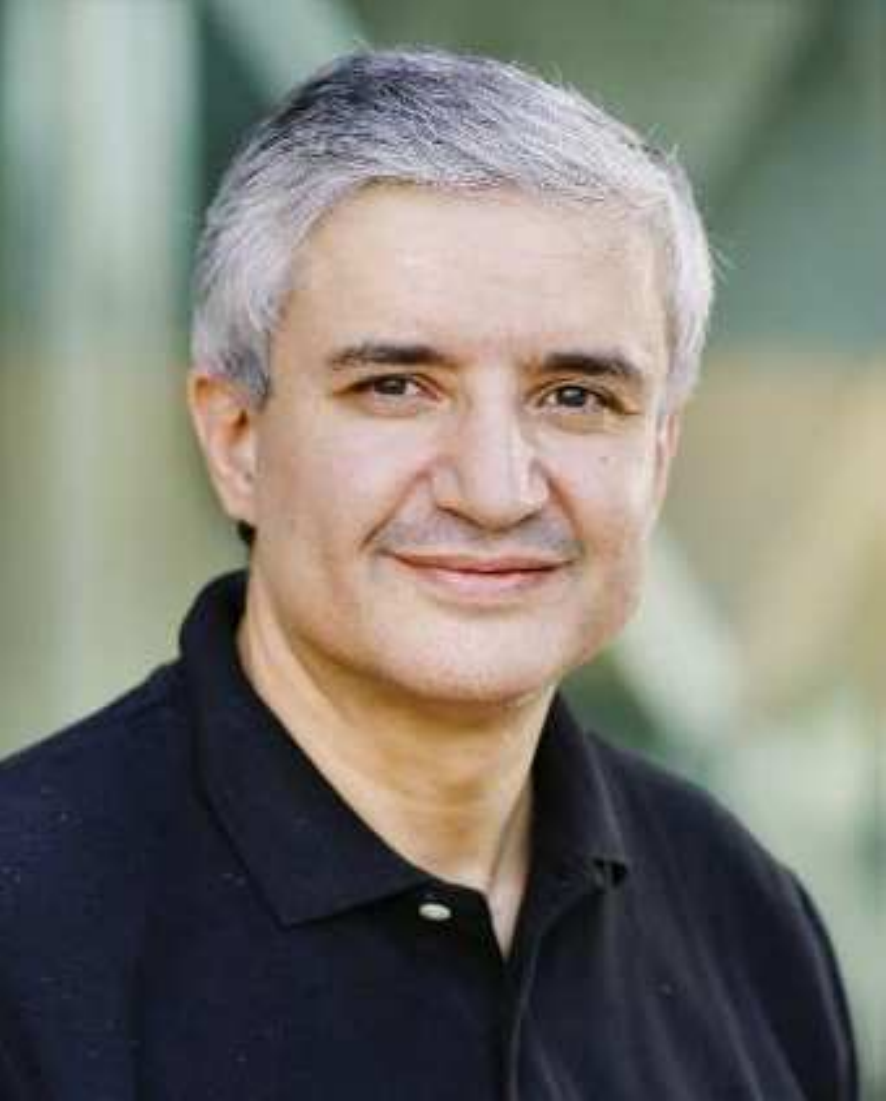}}]{Mohamed-Slim Alouini} (S'94, M'98, SM'03, F'09) was born in Tunis, Tunisia. He received the Ph.D. degree in Electrical Engineering from the California Institute of Technology (Caltech), Pasadena, CA, USA, in 1998. He served as a faculty member in the University of Minnesota, Minneapolis, MN, USA, then in the Texas A$\&$M University at Qatar, Education City, Doha, Qatar before joining King Abdullah University of Science and Technology (KAUST), Thuwal, Makkah Province, Saudi Arabia as a Professor of Electrical Engineering in 2009. His current research interests include the modeling, design, and performance analysis of wireless communication systems.
\end{IEEEbiography}

\begin{IEEEbiography}[{\includegraphics[width=1in,clip,keepaspectratio]{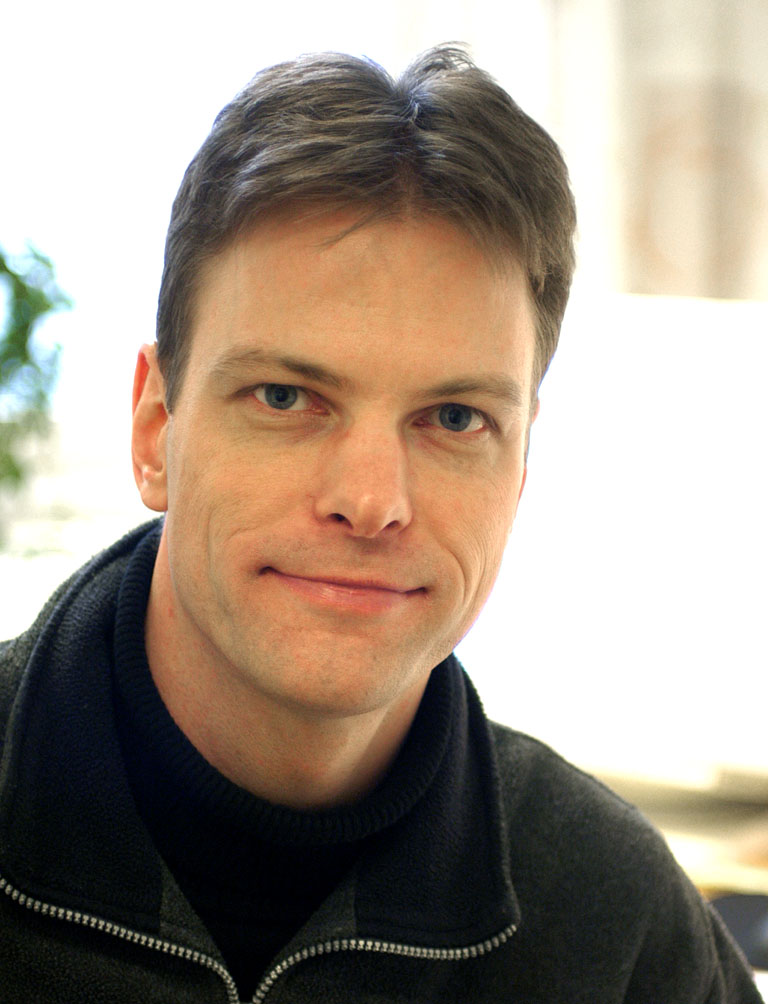}}]{Tommy Svensson} [S’98, M’03, SM’10] is Full Professor in Communication Systems at Chalmers University of Technology in Gothenburg, Sweden, where he is leading the Wireless Systems research on air interface and wireless backhaul networking technologies for future wireless systems. He received a Ph.D. in Information theory from Chalmers in 2003, and he has worked at Ericsson AB with core networks, radio access networks, and microwave transmission products. He was involved in the European WINNER and ARTIST4G projects that made important contributions to the 3GPP LTE standards, the EU FP7 METIS and the EU H2020 5GPPP mmMAGIC and 5GCar projects towards 5G and currently the Hexa-X and RISE-6G projects towards 6G, as well as in the ChaseOn antenna systems excellence center at Chalmers targeting mm-wave and (sub)-THz solutions for 5G/6G access, backhaul/ fronthaul and V2X scenarios. His research interests include design and analysis of physical layer algorithms, multiple access, resource allocation, cooperative systems, moving networks, and satellite networks. \textcolor{black}{He has co-authored 5 books, 100 journal papers, 132 conference papers and 60 public EU projects deliverables. He is Chairman of the IEEE Sweden joint Vehicular Technology/ Communications/ Information Theory Societies chapter, founding editorial board member and editor of IEEE JSAC Series on Machine Learning in Communications and Networks, has been editor of IEEE Transactions on Wireless Communications, IEEE Wireless Communications Letters, Guest editor of several top journals, organized several tutorials and workshops at top IEEE conferences, and served as coordinator of the Communication Engineering Master's Program at Chalmers.
}

\end{IEEEbiography}

\end{document}